\documentstyle[preprint,aps,eqsecnum]{revtex}
\begin{document}
\draft
\title{ Bosonization rules in $1/2 +1$ dimensions}
\author{Manuel Fuentes$^{1,2}$, Ana Lopez$^2$, Eduardo Fradkin$^1$,
and Enrique Moreno$^3$}
\address{$^1$Department of Physics, University of Illinois at
Urbana-\-Champaign , 1110 West Green Street,Urbana, IL 61801-3080 , USA.\\
$^2$ Department of Physics , Theoretical Physics , Oxford University , 1 Keble
Rd. ,
Oxford , OX1 3NP , UK.\\
$^3$Departament d'Estructura i Constituents de la Mat\'{e}ria, Facultat de
F\'{\i}sica,
Universitat de Barcelona, Diagonal ~647, ~08028 Barcelona, Spain,}

\bigskip

\maketitle

\begin{abstract}
We derive the bosonization rules for free fermions on a
half-line with  physically sensible boundary conditions for Luttinger fermions.
We use path-integral methods to calculate the bosonized fermionic currents on
the half-line and derive their commutation relations for a system with a
boundary.
We compute the fermion determinant of the fermionic fluctuations for a system
with a boundary using  Forman's approach.
We find that the degrees of freedom induced at the boundary
do not to modify the commutation relations of the bulk. We give
an explicit derivation of the bosonization rules for the fermion operators for
a
system with boundaries. We derive a set of bosonization rules for the Fermi
operators
which include the explicit effect of the boundaries and of boundary degrees of
freedom.
As a byproduct, we calculate the one-particle Green's function and determine
the effects of the boundaries on its analytic structure.
\end{abstract}
\bigskip


\narrowtext

\def\slp{{\raise.15ex\hbox{$/$}\kern-.57em\hbox{$\partial$}}}
\def\lnA{\raise.15ex\hbox{$/$}\kern-.57em\hbox{$A$}}
\def\slD{\raise.15ex\hbox{$/$}\kern-.57em\hbox{$D$}}
\def\slB{\raise.15ex\hbox{$/$}\kern-.57em\hbox{$B$}}
\def\sls{\raise.15ex\hbox{$/$}\kern-.57em\hbox{$s$}}
\def\slbarA{{\raise.15ex\hbox{$/$}\kern-.57em\hbox{$\bar A$}}}


\section{Introduction}
\label{sec:intro}

The Fermi-Bose equivalence, commonly known as {\it bosonization}, is one of the
most
powerful approaches to extract the non-perturbative behavior of
$1+1$-dimensional
Field theories and strongly interacting Condensed Matter systems. This approach
is an
outgrowth of the work on the solution of the Luttinger-Thirring model by a long
list
of authors~\cite{lieb,luther,coleman,mandelstam,haldane,witten}.
The Fermi-Bose equivalence is deeply rooted in the properties of the algebras
of the
currents and densities of the physical observables. An alternative picture of
the same
physics is also found in the functional integral
approach~\cite{laplata,polyakov}.
In its standard form,
bosonization maps a system of interacting fermions into a system
of bosons. The system of fermions (as well as the equivalent bosonic theory) is
usually defined on manifolds without boundaries. This is appropriate for the
study
of the thermodynamic limit. In the
case of the simple Luttinger-Thirring model, which has an abelian $U(1)$
symmetry, the
bosonic theory is essentially free and the solution of the interacting fermion
theory
is thus found. For more complicated situations, in which the symmetry is
non-abelian,
the bosonized theory is a non-trivial Conformal Field
Theory~\cite{BPZ,KZ,cardy,affleck}.

There are a number of situation of physical interest in which systems with
boundaries
are important. Of particular interest are the so-called quantum wires.
Also, there are a number of situations in which a {\it localized} degree of
freedom
interacts with an extended system of  fermions. Examples are one-dimensional
systems
( as for instance, quantum wires) interacting with magnetic and non-magnetic
impurities.
The interplay of Luttinger-like behavior and the Kondo effect is a subject of
considerable interest~\cite{meir,kane}. Similar type of physics is also found
in
the context of the Callan-Rubakov effect~\cite{callan}. A number of authors
have
used bosonization methods to the study of semi-infinite systems~\cite{kane}
 and systems with quantum impurities~\cite{affleckimp} such as the Kondo
effect~\cite{Affleck} and spin chains with impurities~\cite{affleckspin}.
However, in all cases, the
bosonization rules of the {\it bulk} system were used and little attention was
paid to
the role that the boundary conditions of the fermions may play in the bosonized
theory.

In this paper we discuss the effects of the boundary conditions of the fermions
and
in which way they affect the physics of the bosonized theory. Since
bosonization is
seen most clearly in the case of free fermions, we consider here just the
simplest case, namely non-interacting, abelian, theory of  fermions on a
half-line
(``${\frac{1}{2}}+1$-dimensions"). In a
separate publication~\cite{enviasdedesarrollo} we will discuss the case of
interactions and the Kondo effect, which involves systems with non-abelian
symmetries.
We consider systems of fermions with two types of boundary conditions: (a)
$R=-L$ {\it
at} the boundary and (b) $R(0,x_2)=-e^{2i\theta(x_2)} L(0,x_2)$, again {\it at}
the
boundary. This choice of boundary conditions is motivated by the underlying
microscopic
physics of the systems of interest. In the case of quantum wires, the
microscopic system consists of non-relativistic fermions at finite density. At
a
physical edge ({\it i.~e.\/} the boundary) the fermion amplitude $\Psi(0)$ must
vanish. Upon a decomposition into Left and Right movers, we find that a sharp
edge
implies the choice (a), {\it i.~e.\/} $R(0)=-L(0)$. If a localized quantum
impurity,
parametrized by a degree of freedom $\theta$, is
placed at the edge, the boundary condition changes to case (b),{\it i.~e.\/}
$R(0,x_2)=-e^{2i\theta(x_2)} L(0,x_2)$. Physically, this boundary condition is
caused
by a coupling between the fermion charge density at the edge and the localized
quantum
impurity. We also introduce physically sensible models of the quantum dynamics
of the
impurity.

In section \ref{sec:Fermions}, we study a system of free electrons confined
to the half-line and derive an expression for its partition function.
In order to write this partition function only in terms
of bosonic operators, in Section \ref{sec:borde} we compute the determinant of
the $1+{1\over{2}}$-dimensional Dirac operator coupled to a source which is
defined
as a delta function at the boundary. We show that this problem is equivalent to
the
computation of the determinant of the Dirac operator plus a suitable boundary
condition . This last determinant is solved by using Forman's
approach\cite{Forman}.

In section \ref{sec:currents}, we derive an expression for the partition
function only in terms of bosonic degrees of
freedom. In other terms, we present an alternative form of
abelian bosonization in terms of functional integrals. Although
this is a subject that has been discussed extensiveley in the
literature, we present here a derivation of this classic result in
a form that is suitable for treating systems with boundaries. We use
these results to calculate the bosonized currents, the current correlation
functions
and their anti-commutation relations paying close attention to the
role  of the boundary conditions of the fermions.

In section \ref{sec:bosofermiops}, we derive the one particle fermionic Green's
function from the bosonized theory. We discuss two cases, (a) when the boundary
condition is $R=-L$ ,
and (b) when there is an extra degree of freedom at the boundary, {\it
i.~e.\/},
$R(0,x_2)=-e^{2i\theta(x_2)} L(0,x_2)$. We focus on the way the one particle
Green's function is affected by the presence of the boundary and of a dynamical
degree
of freedom at the boundary. In section~\ref{sec:dynamical} we discuss  a model
of
free fermions coupled to a dynamical boundary degree of freedom, and
calculate the fermion one-particle Green's function for this case of
interest. We consider two asymptotic regimes: (a) one in which the
boundary degree of freedom is strongly pinned, and (b) the case in which it
fluctuates
wildly. In the Conclusions we give a brief summary of our results.
In the appendices we give details of
the computation of the boundary contributions to the fermion Jacobian,
of the use of Forman's method and to Euclidean-Minkowski
correspondences are given in the Appendices.

\section{Free Fermions on a half-line}
\label{sec:Fermions}

In this section we study a system of  free massless Dirac fermions confined to
the half
line ({\it i.\ e.\ }, $1+{1\over{2}}$ dimensions) satisfying dynamical boundary
conditions. Our goal is to compute the partition function in order to be able
to
derive the bosonizations rules for this system. Throughout, we work in
Euclidean space. In practice, this means that the space-time manifold is a
semi-infinite long
cylinder, with its axis along the space direction $x_1$ and periodic
(antiperiodic for fermi
fields) boundary conditions along the imaginary time direction $x_2$ of
perimeter $T$, with $T \to \infty$.

The (Euclidean) Lagrangian for the system (coupled to sources) is
\begin{equation}
{\cal L}_{\rm F}={\bar \psi} \;i  \slp \; \psi+{A}_\mu J_\mu
\label{eq:Lth}
\end{equation}
where ${A}_\mu$ is a source (a background gauge field) which
couples to the fermion current $J_\mu={\bar \psi}\; \gamma_\mu \psi $.
In a Fermi theory in Euclidean space, not just the time coordinate has to be
continued analytically to the imaginary axis. The structure of the
$\gamma$-matrices and the requirement that the Euclidean Dirac operator be
hermitean forces an analytic continuation of the chiral angle. Hence,
the chiral transformations  generated by $\gamma_5$ become complexified in
Euclidean
space.
However, we must keep in mind that these
two analytic continuations are logically distinct. Moreover, these two analytic
continuations need to be done in
order to have a physically sensible interpretation of the results and to
recover results back in Minkowski space.
Hence, the fermion boundary conditions themselves also have to be continued
analytically. We thus demand that the fermions satisfy the boundary condition
 $R(0,x_2) = - e^{2\theta(x_2)} L(0,x_2)$.

The dynamics of the fermions is described by the functional integral
\begin{equation}
{\cal Z}[A]= \int {\cal D} {\bar \psi} {\cal D} \psi
\exp \left(-\int d^2x {\cal L}_F \right).
\label{eq:functint}
\end{equation}

In what follows we will use the path integral bosonization
method based on Seeley's expansions for complex powers of elliptic
operators\cite{seeley}.

The first step in the bosonization process is to decouple the
fermions from the sources by means of a combination of (suitably chosen) gauge
and chiral
smooth, single-valued transformations of the form
\begin{eqnarray}
\psi(x)&=&e^{i\eta(x)+\gamma_5 \phi(x)}\chi(x)\nonumber\\
{\bar \psi}(x)&=&{\bar \chi}(x) \; e^{-i\eta(x)+\gamma_5 \phi(x)},
\label{eq:chiral}
\end{eqnarray}
provided that the vector potentials $A_\mu(x)$ can be written in the
form
\begin{equation}
A_\mu(x)=\partial_\mu \eta(x)-\epsilon_{\mu \nu} \partial{_\nu}\phi(x).
\label{eq:vecpot}
\end{equation}
This requirement is satisfied by all topologically trivial
configurations of the fields $A_\mu$.
The usefulness of the transformations of Eq.~(\ref{eq:chiral}) and
Eq.~(\ref{eq:vecpot}) is that they completely decouple the fermions
from the vector potentials.
However, this transformation changes the fermionic measure in a non-trivial
way.
Namely,
the change in the fermionic measure under this transformation is:
\begin{equation}
 {\cal D}{\bar \psi} \; {\cal D}{\psi} \; = J_F {\cal D}{\bar \chi} \;
 {\cal D}{\chi}. \;
\end{equation}
The jacobian, $J_F$ , can be written as
\begin{eqnarray}
J_F = {{\rm Det} ( i \slp + \lnA)\over
{\rm Det} ( i \slp)} = e^{- \int _0^1 w^\prime (t) dt}\;
\end{eqnarray}
with
\begin{equation}
w^\prime = 2 \int dx_2 dx_1 tr K_2[i\slD_t;x,x] \gamma^5 \phi (x)
\end{equation}
where $i\slD_t =i\slD_t[(1-t)A]$. The kernel $K_2[i\slD_t;x,x]$ is the
constant
term in the analytic expansion of the Heat Kernel $< x | e^{- s\slD_t
 \slD_t} | x>$.
In the case of the fermions on the full line (which can be considered as a
close
manifold if we impose vanishing boundary conditions at the infinity) the
jacobian is
well known and it has been calculated by several authors\cite{laplata}. In the
case
of a
manifold with boundary we have to evaluate this Heat Kernel, following
Atiyah-Patodi-Singer (APS)\cite{APS}, separating the contributions of the
boundary
from the ones of the bulk. Hence, the decoupling of the gauge field has to be
done in
two steps, first we
 decouple the bulk and then the boundary.
For this pourpose it is convenient to write the field $A_\mu$ as $A_\mu=
B_\mu+s_\mu$
where $s_\mu$ is defined as the value of  $A_\mu$ at the boundary, and $B_\mu$
as its
value in the bulk. That is, $s_\mu=A_\mu \delta(x_1)$ and $B_\mu$ is zero at
the
origen.

We define a chiral and a gauge transformation as in Eq.~(\ref{eq:chiral}) ,
such
that
it only decouples the field
$B_\mu$ from the fermions. Therefore  at this point we are going to assume that
the
field $B_\mu$ is the one that
can be written in terms of $\phi$ and $\eta$ as in Eq.~(\ref{eq:vecpot}) .
The partition function becomes
\begin{equation}
{\cal Z}[B_\mu,s_\mu]=\int {\cal D} {\bar \chi} {\cal D} \chi \; J_F \;
\exp \left( - \int_\Omega d^2x [{\bar \chi} (i \slp+\sls)\chi]\right)
\label{eq:functint3}
\end{equation}
where the integral in the action is restricted to the half-plane
$\Omega=\{\vec{x}|\; x_1\geq0\}$.
The Jacobian reads,
\begin{equation}
J_F = {{\rm Det}( i \slp + \slB+ \sls)\over
{\rm Det} ( i \slp +\sls)} = e^{- \int _0^1 w^\prime (t) dt}.\;
\end{equation}
The APS procedure to compute the Heat Kernel consists in dividing the domain
($\Omega$) in two intervals, $(0,\epsilon)$ and $(\epsilon,\infty)$.
In the interval $(\epsilon,\infty)$ we can compute $w'(t)$ in a usual
way\cite{laplata} and obtain the well-known result for closed compact
manifolds
\begin{equation}
J_F = e^{ -{1\over{2\pi}}\int_{\Omega(\epsilon)} (\partial_{\mu}\phi)^2 +
{1\over{\pi}}\int_{\Omega(\epsilon)} {\epsilon}_{\mu\nu}s\mu
\partial_{\nu}\phi }
\label{omegaepsilon}
\end{equation}
where $\Omega(\epsilon)=\{\vec{x}|\; x_1>\epsilon\}$.
The contribution of the boundary to the jacobian is given by the regular term
of the following expresion, (in the limit $s\rightarrow0$)
\begin{equation}
{\delta}J_{F}=\exp \left( \int_0^1 dt\;\int dx_2\;\int^\epsilon_0 dx_1< x |
e^{-s \slD_t \slD_t} |
x>
\phi(x_1,x_2) \right).
\end{equation}
We show in Appendix A that this contribution is given by
\begin{equation}
{\delta}J_{F} =\exp\left( -{1\over{4\epsilon}}\int dx_2 \phi(0,x_2)\right)
\end{equation}
Therefore, if we choose $\phi (0,x_2)=0$ the contribution of the boundary term
to the
jacobian is one. It is important to remark
that this cancellation happens only because we made a chiral transformation
that
is trivial at the boundary. The price for this choice is that we still have the
fermions coupled to the field $s_\mu$ with support only on the boundary.
We will show how to deal with this problem in the next section.

Then after the chiral transformation is performed the partition function is
\begin{equation}
{\cal Z}[B_\mu,s_\mu]=\int {\cal D} {\bar \chi} {\cal D} \chi \;
\exp \left( - \int_\Omega d^2x [{\bar \chi} (i \slp+\not\!s)\chi]\right)
\exp \left(-{1\over{2\pi}}\int_\Omega (\partial_{\mu}\phi)^2 + {1\over{\pi}}
\int_\Omega {\epsilon}_{\mu\nu}s_\mu
\partial_{\nu}\phi \right)
\label{eq:functint2}
\end{equation}
where we took the limit $\epsilon\rightarrow0$.

Since
\begin{equation}
B_\mu= -(\epsilon_{\mu\nu}\partial_{\nu}\phi-\partial_{\mu}\eta).
 \label{eq:u,v}
\end{equation}
we can write
\begin{equation}
\partial_{\nu}\partial_{\nu}\phi(x)=\epsilon_{\alpha\mu}\partial_{\alpha}B_{\mu}(x)
\end{equation}
with the boundary condition
\begin{equation}
\phi(0,x_2)=0
\end{equation}
Therefore
\begin{equation}
\phi(x)=\int_\Omega d^2y \;G_0(x,y)
\epsilon_{\alpha\mu}\partial^{(y)}_{\alpha}B_\mu(y)
\end{equation}
where $G_0(x,y)$ is the Green's funcion defined as
\begin{equation}
 \left\{
\begin{array}{ll}
\partial ^2_x G_0(x,y)= \delta(x-y)\\
G_0(0,x_2;y)=0 .
\end{array}
\right.
\label {eq:G0}
\end{equation}
Since $B_{\mu}$ is only defined for $x_1> 0$, we have
 to use the Green's function
for the half line (Eq.~(\ref{eq:G0})) which is given by
\begin{equation}
G_0(x,y)= g(x_2-y_2,x_1-y_1) - g(x_2-y_2,x_1+y_1)
\label{eq:solution}
\end{equation}
where
 \begin{equation}
g(u,v) = {1\over {4\pi}} \ln {{(u^2 + v^2 + a^2)} \over{a^2}}
\label{g0}
\end{equation}
with $a$ acting as a regulator.

Using these expressions to write $\phi(x)$ in terms of $B_{\mu}$ , the jacobian
$J_F$
can be written as
\begin{eqnarray}
ln J_F =
&&-{1\over{2\pi}}\int_\Omega (\partial_{\mu}\phi)^2 +
{1\over{\pi}}\int_\Omega {\epsilon}_{\mu\nu}s_\mu
\partial_{\nu}\phi \nonumber\\
&&={1\over{2\pi}}\int_\Omega dx  \int_\Omega dy\left( B_{\mu}(x)
\Gamma_{\mu\nu}(x,y)
B_{\nu}(y) + 2 B_{\mu}(x) \Gamma_{\mu\nu}(x,y) s_{\nu}(y) \right)\nonumber\\
&&={1\over{2\pi}}\int_\Omega dx  \int_\Omega dy (B_{\mu}(x)+ s_{\mu}(x))
\Gamma_{\mu\nu}(x,y)
 (B_{\nu}(x)+ s_{\nu}(x)) \nonumber\\
&&-{1\over{2\pi}}\int_\Omega dx  \int_\Omega dy s_{\mu}(x) \Gamma_{\mu\nu}(x,y)
s_{\nu}(y)
\label{eq:functionU2}
\end{eqnarray}
where  ${\Gamma}_{\mu \nu}(x,y)$ is defined as
\begin{equation}
{\Gamma}_{\mu \nu}(x,y)=
\left(\partial^{(x)}_{\alpha}\partial^{(y)}_{\alpha}\delta_{\mu \nu}-
\partial^{(y)}_{\mu}\partial^{(x)}_{\nu}\right)G_0(x,y)
\label {eq:gammas}
\end{equation}

We can linearize the cuadratic term in $B_{\mu}(x)+ s_{\mu}(x)$ by introducing
a
 bosonic
field $\omega$ defined in  $\Omega$ such that $\omega(0,x_2)=0$. It is easy to
see that
Eq.~(\ref{eq:functionU2}) can be expressed as
\begin{eqnarray}
\ln J_F &=&{\cal K} \int  \;{\cal D} \omega \exp \left(
-{1\over{2\pi}}\int_\Omega
(\partial_{\mu}\omega)^2
+ {i\over{\pi}}\int_\Omega
{\epsilon}_{\mu\nu}\partial_{\nu}\omega(s_\mu+B_\mu)\right)\nonumber\\
&&\exp \left(-{1\over{2\pi}}\int_\Omega dx  \int_\Omega dy s_{\mu}(x)
\Gamma_{\mu\nu}(x,y) s_{\nu}(y)\right)
\end{eqnarray}
Since by definition, $s_\mu(x)= s_\mu(x_2) \delta(x_1)$, we need the value of
${\Gamma}(x,y)$ at $x_1=y_1=0$. It is easy to check that
${\Gamma}_{12}|_{x_1,y_1=0}={\Gamma}_{21}|_{x_1,y_1=0}=0$. Similarly, by using
the
expression for the Green's function, Eqs.~(\ref{eq:solution}-\ref{eq:G0}), and
for the
kernels $\Gamma_{\mu \nu}(x,y)$, Eq.~(\ref{eq:gammas}), we also find that,
in the limit $a\rightarrow0$, ${\Gamma}_{11}|_{x_1,y_1=0}=0$.
In the same way, we find that the value of
${\Gamma}_{22}(x,y)|_{x_1,y_1=0}$ is given by
\begin{equation}
{\Gamma}_{22}(x,y)|_{x_1=y_1=0}= -{1\over{\pi}}{\cal P} \frac
{1}{(x_2-y_2)^2+a^2}
+ {1\over{a}}\delta(x_2-y_2)
\end{equation}
where $\cal P$ denotes the principal value. From now on we will assume a
principal value in this type of expressions.
Then the partition function becomes
\begin{eqnarray}
{\cal Z}[b_\mu,s_\mu]&=&{\cal K} \int {\cal D} {\bar \chi} {\cal D} \chi \;
{\cal D} \omega
\;\exp \left( - \int_\Omega d^2x [{\bar \chi} (i
\slp+\not\!s)\chi]\right)\nonumber\\
&&\exp \left(-{1\over{2\pi}}\int_\Omega  (\partial_{\mu}\omega)^2
+ {i\over{\pi}}\int_\Omega
{\epsilon}_{\mu\nu}\partial_{\nu}\omega(s_\mu+B_\mu)\right)
\nonumber\\
&&\exp\left(-{1\over{4\pi^2}} \int \frac
{(s_2(y_2)-s_2(x_2))^2}{(x_2-y_2)^2}\right).
\label{pf}
\end{eqnarray}
Now we have solved completely the decoupling of the gauge field $B_\mu$.
But we still have a fermionic integration left to perform.
This integration will be done in the next section.

\section{Free Fermions in a $\delta(x_1)$-potential}
\label{sec:borde}

In this section we calculate  the determinant of the
$(1+{1\over{2}})$-dimensional Dirac operator coupled to a $\delta(x_1)$
potential.
Hence we have to consider the Dirac operator,
$[\gamma^\mu(i\partial_\mu+s_{\mu})]$,
where
$s_\mu = s_\mu(x_2)\delta(x_1)$.
The formal way to compute the determinant of this operator is by computing
its eigenvalues. First, it is useful to regularize the $\delta$ function as
the limit $\epsilon \rightarrow 0$ of the function $v_{\epsilon}(x)$ where
$\int_{\epsilon - {\delta \over 2}}^ {\epsilon + {\delta \over 2}}
v_{\epsilon}(x)=1$.
It is clear that when $\epsilon,\delta \rightarrow 0$, the solutions of
\begin{equation}
[{\gamma}^\mu  ( i \partial_\mu + s_{\mu})]\psi=\lambda \psi
\label {delta}
\end{equation}
in $x_1\geq0$ will be the same as the solutions of the equation
\begin{equation}
{\gamma}^\mu  ( i \partial_\mu )\psi=\lambda \psi
\end{equation}
plus a suitable chosen boundary condition.
To find such a boundary condition we observe that
integrating Eq.(\ref{delta}) over the interval $(\epsilon - {\delta \over 2},
\epsilon + {\delta \over 2})$ and assuming that the field $\psi$ is finite
in such interval, we obtain for small $\delta$
\begin{equation}
-{\gamma}_1 (\psi (\epsilon^{+})-\psi (\epsilon^{-})) +
i \left({\gamma}_1 s_1(x_2) + {\gamma}_2 s_2(x_2) \right)
\int_{\epsilon - {\delta \over 2}}^{\epsilon + {\delta \over 2}} dx_1\;
\psi(x_1)
v_{\epsilon}(x_1) = 0.
\end{equation}
Such a field can be found if we look at solutions of
\begin{equation}
\partial_1\psi =\left( {\gamma}_5s_2(x_2) + i s_1(x_2) \right)
v_{\epsilon}(x_1)\psi(x_1).
\end{equation}
Explicitly,
\begin{equation}
\psi (x)= {\hat P}_{x_1} \exp[ \left( {\gamma}_5 s_2(x_2)
 + i s_1(x_2) \right) \int_{{\epsilon}^{-}}
^{x_1} dyv_{\epsilon}(y)  ] \psi({\epsilon}^{-})
\label{order}
\end{equation}
where ${\hat P}_{x_1}$ is the spatial ordering operator. Since the
operator in the exponent of Eq.~(\ref{order})
commutes at spatially separated points because of the shape of
$v_{\epsilon}(x)$, we may set  ${\hat P}_{x}=1$, and taking the limit
$\delta \rightarrow 0$ we get
\begin{equation}
\psi({\epsilon}^{+})= e^{{\gamma}_5s_2(x_2) +i s_1(x_2)}
\psi({\epsilon}^{-}).
\end{equation}
It is clear now that, in the limit $\epsilon \rightarrow 0$, the boundary
condition we are looking for becomes
\begin{equation}
\psi({\epsilon}^{+})= e^{-{\gamma}_5s_2 (x_2)+i s_1(x_2)}
\psi({\epsilon}^{-}).
\end{equation}
Recalling that at $\epsilon^-$ the fields also satisfy $Be^{-
\gamma_5\theta(x_2)}\psi(\epsilon^-)=0$ with B the matrix given by
\begin{eqnarray}
B \: &=&
\: \left(
\begin{array}{cc}
1 & 1 \\
1 & 1
\end{array} \; \right)\;
\end{eqnarray}
 then, in the limit $\epsilon \rightarrow0$, the boundary condition for the
field $\psi$ is
\begin{equation}
B{\cal U}^{-1}(x_2)\psi|_{x_1=0}=0
\label{bc}
\end{equation}
where ${\cal U}(x_2)= e^{[{\gamma}_5(\theta(x_2) + s_2(x_2))]}$.
This equation tells us that the source $s_2(x_2)$ can be though as a dynamical
degree
of freedom at the boundary that interacts with the fermions through the above
boundary condition.
In other words, the determinant we have to compute is
$det {({\gamma}^\mu  ( i \partial_\mu ))}_{B{\cal U}^{-1}}$
where
\begin{equation}
{{\gamma}^\mu  ( i \partial_\mu )}_{B{\cal U}^{-1}}= \left\{
\begin{array}{ll}
{\gamma}^\mu  ( i \partial_\mu ) \;{\rm{ with}\; \psi \;{\rm such\; that}}\\
{B{\cal U}^{-1}(x_2)\psi|_{x_1=0}=0} .
\end{array}
\right.
\label{opconborde}
\end{equation}
The solution for this kind of determinants was first given by
Forman\cite{Forman}.
His theorem
relates the determinant of the above differential operator with the determinant
of an (infinity dimensional) matrix,
\begin{equation}
{\cal D}et {({\gamma}^\mu  ( i \partial_\mu ))}_{B{\cal U}^{-1}(x=0)}{\sim}
\int ds\;Tr({d\over ds}H(s) H^{-1}(s))
\label{cebo}
\end{equation}
The matrix $H$ is a functional of the function $\theta(x_2) + s_2(x_2)$.

 In order to apply Forman's methods we introduce an auxiliar parameter $\tau$
such that
\begin{equation}
{\cal U}(\tau)= e^{\tau {\gamma}_5 \alpha}
\end{equation}
where we have defined $\alpha(x_2)= \theta(x_2) + s_2(x_2)$.
This method relies on the knowledge of the space of solutions of the
homogeneous
equation $( i \gamma_\mu \partial_\mu )\psi=0$. We must consider a complete,
but
otherwise arbitrary, system of functions in the kernel of this differential
operator, which do not need to satisfy any particular boundary condition.
In order to simplify the calculations, we choose the basis that expands
the
kernel of $( i \gamma_\mu \partial_\mu )\psi=0$, satisfying
\begin{eqnarray}
\psi_{n}(x_1,-{T\over {2}})&=& -\psi_{n}(x_1,{T\over {2}})\;,\nonumber \\
B\psi_{n}(x_2,0)&=&B\psi_{n}(x_2,L)=0\;.
\end{eqnarray}
Later on we will send $L,T\rightarrow \infty$ .
The basis is
\begin{eqnarray}
\psi_{n}(x_2,x_1)=e^ {iw_{n}[x_2-i\gamma_{5}(x_1-L)]}\left(
\begin{array}{cc}
1 \\
1
\end{array} \;\; \right)\;
,\; n\in \cal{Z}
\end{eqnarray}
where $w_{n}= {{(2n+1)\pi}\over {T}}$.
Forman's Theorem \cite{Forman} relates the problem of evaluating the
functional determinant of a differential operator with a parameter-dependent
boundary condition (such as the one defined by Eq.~(\ref{opconborde})), to that
of
obtaining the determinat of an operator $\Phi_{\mu}(\tau)$, acting on functions
defined on the boundary :
\begin{equation}
{d\over {d\mu}}ln\;\left({
{\rm Det} ( i \slp )_{B{\cal U}^{-1}(\mu + \tau)}}
\over {{\rm Det} ( i \slp)_{B{\cal U}^{-1}(\mu)}}
\right) =
{d\over {d\mu}}ln\; \det \Phi_{\mu}(\tau).
\label{forman}
\end{equation}
 Here the determinants on the left-hand side are defined through the $\zeta$
-function regularization, while the right-hand side is a well-defined quantity
that can be evaluated using any basis at the boundary. The operator
$\Phi_{\mu}(\tau)$ is defined as follows. We first define the Poisson map
$P_B$. If the boundary conditions are defined with the operator $B$, then $P_B$
is such that the $unique$ solution of
\begin{equation}
 \left\{
\begin{array}{ll}
{\cal O}f(x) = 0 \; x\in \Omega\\
Bf(x)= h(x)\; x\in \partial \Omega
\end{array}
\right.
\end{equation}
where ${\cal O}$ is an arbitrary differential operator, is given by
$f(x)=P_B h(x)$. If $A$ is another boundary condition, then
$\Phi_{AB}=A P_B$. Even though $P_B$ is difficult to evaluate, $\Phi_{AB}$ is
not because of its main property
\begin{equation}
A \psi= \Phi_{AB} B\psi
\end{equation}
 for any solution of ${\cal O} \psi =0$.
In our case $\Phi_{\mu}(\tau)$ satisfies
\begin{equation}
B {\cal U} ^{-1} (\mu + \tau) \psi =
                  \Phi_{\mu}(\tau) B {\cal U} ^{-1} (\mu) \psi
\end{equation}
for any $\psi $ that obeys the free Dirac equation. Let us define
$h_{\mu}^{n} (x_2)$ as
\begin{equation}
h_{\mu}^{n} (x_2)= B {\cal U} ^{-1} (\mu) \psi_{n}(x_2)
\end{equation}
where $\psi_{n}(x_2)= \psi_{n}(0,x_2)$.
They satisfy  $h_{\mu + \tau}^{n} (x_2) = \Phi_{\mu}(\tau) h_{\mu}^{n} (x_2)$.
If we were able to write $h_{\mu}^{n} (x_2)$ and $h_{\mu + \tau}^{n} (x_2)$
in terms of
a complete set of functions in $[0,L]$ we would have the (infinite) matrix
$\Phi_{\mu}(\tau)$ expressed in this basis, and by using conventional methods
we
could compute its determinant.
The basis will be $ \psi_{n}(L,x_2)= e^{ iw_{n} x_2}$ with
$w_{n}= {{(2n+1)\pi}\over {T}}$
$( n\in \cal Z)$. It is easy to compute $h_{\mu}^{n}(x_2)$ and
$h_{\mu + \tau}^{n} (x_2)$
in such basis. Following the definition,
\begin{eqnarray}
h_{\mu}^{n} (x_2)&=& \cosh (w_{n}L + {\mu}\alpha) e^ { iw_{n} x_2}\;,\nonumber
\\
h_{\mu + \tau}^{n}(x_2)&=& \cosh (w_{n}L + (\mu + \tau )\alpha) e^{iw_{n}
x_2}\;.
\end{eqnarray}
In the limit $L\rightarrow \infty$, $\tanh (w_{n}L)$ goes to ${\rm
sign}(w_{n})$, where
\begin{equation}
{\rm sign}(w_{n})= {\rm sign} (2n+1) =\left\{
\begin{array}{ll}
\;1, \; n \geq 0 \\
-1,\; n < 0
\end{array}
\right.
\end{equation}
Then, up to a normalization constant, we can write
\begin{equation}
h_{\mu}^{n} (x_2) = e^{{\rm sign}(w_{n}) \mu \alpha(x_2)}e^{ iw_{n} x_2}
\end{equation}
Note that the m-th component of $h_{\mu}^{n} (x_2)$ in the basis under
consideration is
\begin{equation}
h_{\mu}^{n,m} (x_2) = {1\over {2T}} \int ^{T\over {2}}_{-T\over {2}}  dx_2 \;
e^ {[ i(w_{n}-w_{m}) x_2]} \;e^ {[{\rm sign}(w_{m}) \mu \alpha(x_2)]}
\end{equation}
We can think of $h_{\mu}^{n,m} (x_2)$ as matrix elements, in the space spanned
by
the basis $\psi_{n}(0,x_2)$, of a matrix $ h(\tau)$. In this way,
\begin{equation}
<\psi_{n}|h(\tau)|\psi_{m}> = h_{\mu}^{n,m}
\end{equation}
and the operator $\Phi_{\mu}(\tau)$ can be written as
\begin{equation}
h(\mu + \tau) h^{-1}(\tau)= \Phi_{\mu}(\tau)
\end{equation}

Then
\begin{equation}
{d\over {d\mu}} \ln\; \det \Phi_{\mu}(\tau)=
Tr
{d\over {d\mu}}[ h( \mu+\tau)h^{-1}(\mu+\tau)] -
{d\over {d\mu}}[ h( \mu)h^{-1}(\mu]) ].
\end{equation}
We have reduced the computation of ${d\over {d\mu}}ln\; Det \Phi_{\mu}(\tau)$
to the computation of a quantity of
the form
\begin{equation}
Tr[{d\over{d\mu}}h(\mu) h^{-1}(\mu)]
\end{equation}
at $\mu=1$, where $\mu$ is an arbitrary parameter, and $h(\mu)$ is a matrix of
the form
\begin{equation}
h(\mu) \: =
\: \left(
\begin{array}{cc}
{[e^{-\alpha}]}_{--} & {[e^{\alpha}]}_{-+} \\
{[e^{-\alpha}]}_{+-} & {[e^{\alpha}]}_{++}
\end{array} \;\; \right)\;
\end{equation}
(see Appendix B for notation). At this point it is important to make a
conection between the matrix $h(\mu)$ and the so-called
Toeplitz matrices. A Toeplitz matrix of order $N$ is the name for a matrix
$T^{(N)}$ of the form
\begin{equation}
T^{(N)}_{m,n}= c_{m-n}\;\;m,n = 0,....,N-1
\end{equation}
where $c_{p}$ are arbitrary complex numbers $(p=0,\pm 1,....,\pm (N-1))$.
Let us denote $D_{N}$ as its determinant.
Giving a set of complex numbers $c_{p},\;p\geq 0$, and under suitable
conditions
{\it Szego's Theorem} \cite{cebo} gives an expresion for
\begin{equation}
\lim_{N\rightarrow \infty}({D_{N}\over{{\mu}^N}})
\end{equation}
Clearly ${{\mu}^N}$ acts as a regulator.
If
\begin{equation}
c_p = {1\over {2\pi}} \int _0^{2\pi} dt \;e^{-ipt}\;C(e^{it})
\end{equation}
and
\begin{equation}
\mu= \exp\{{1\over {2\pi}}\int _0^{2\pi} dt\;C(e^{it})\}
\end{equation}
and
\begin{equation}
g_p= {1\over {2\pi}} \int _0^{2\pi} dt \;e^{-ipt}\;lnC(e^{it})
\end{equation}
then, {\it Szego's Theorem} states that
\begin{equation}
\lim_{N\rightarrow \infty}({D_{N}\over{{\mu}^N}})=
\exp\{\sum_{p=1}^{\infty} \;p g_{-p}g_p\}.
\end{equation}
It is obvious that ${[e^{-\alpha}]}_{--}$ and ${[e^{\alpha}]}_{++}$ are both
Toeplitz blocks in the sense that they satisfy the definition of a Toeplitz
matrix. Notice that if $h(\mu)$ were made up by only one of these blocks, the
determinant (6.19) would be exactly the one given by  $Szego's \;Theorem$.
We are going to show  that, even though the off-diagonal blocks are
non-vanishing, and the indices that label the matrix entries are
integers numbers, $Szego's \;Theorem$ is still valid up to an overall factor of
2.
We can also see that Forman's Theorem provides a natural cancelation for
the regularization factors in the Szego's Theorem.

In Appendix B we show that
\begin{eqnarray}
Tr[{d\over{d\mu}}h(\mu) h^{-1}(\mu)] &=&
tr \left\{ [\alpha]_{++} -  [\alpha]_{--} \right.\nonumber\\
&+& [\alpha]_{+-}( -{[e^{{\alpha}_-}]}_{--} {[e^{{-\alpha}_{-}}]}_{-+} +
{[e^{{\alpha}_{-}}]}_{--}{[e^{{\alpha}_{-}}]}_{-+}{[e^{{-\alpha}_{-}}]}_{++}
{[e^{{-\alpha}_{-}}]}_{++} ) \nonumber\\
&+&\left.
[\alpha]_{-+}({[e^{{-\alpha}_+}]}_{++}{[e^{{\alpha}_+}]}_{+-} -
{[e^{{-\alpha}_{+}}]}_{++}{[e^{{-\alpha}_{+}}]}_{+-} {[e^{{\alpha}_{+}}]}_{--}
{[e^{{\alpha}_{+}}]}_{--}) \right\}
\label{traza}
\end{eqnarray}

After anti-transform Fourier the above expression, we see that the first two
terms
cancell with each other. The third term becomes
\begin{eqnarray}
Tr\{[\alpha]_{+-}{[e^{{\alpha}_-}]}_{--} {[e^{{-\alpha}_-}]}_{-+}\}&=&
\sum _{n\geq0p,q<0}[\alpha]_{n-p}{[e^{{\alpha}_-}]}_{p-q}
                                       {[e^{{-\alpha}_{-}}]}_{q-n} \nonumber\\
&=&\sum _{n\geq0}\sum _{p>0}\sum_{q=1}^{p-1}
[\alpha]_{n+p}{[e^{{\alpha}_-}]}_{-p+q}{[e^{{-\alpha}_-}]}_{-q-n} \nonumber\\
&=&\oint {[dz_i]\over{\prod_{i}z_i}}\sum _{n\geq0}\sum _{p>0}\sum_{q=1}^{p-1}
\alpha_{+}(z_1)
e^{{\alpha}_{-}(z_2)}e^{{-\alpha}_{-}(z_3)}z_1^{-(n+p)}z_2^{(p-q)}
z_3^{q+n} \nonumber\\
&=&- \oint [dz_1]\alpha(z_1)\;\oint
[dz_2]{e^{{\alpha}_{-}(z_2)}\over{(z_2-z_1)}}
\;\oint [dz_3]{e^{{-\alpha}_{-}(z_3)}\over (z_3-z_1)^2} \nonumber\\
&=&-\oint
[dz_1]\alpha_{+}(z_1)e^{{\alpha}_{-}(z_2)}(-\partial_{z_1}{\alpha}_{-}(z_1))
 e^{-{\alpha}_{-}(z_2)}\nonumber\\
&=&{1\over {2{\pi}i}}
\int dx_2 \alpha_{+}(x_2)\partial_{2}{\alpha}_{-}(x_2).
\end{eqnarray}

 Folowing the same procedure it is easy to see that the fourth term gives
exactly
the same contribution. The two terms with $[\alpha]_{-+}$ contribute with the
same kind of expresion but with $``+"$ interchanged with $``-"$ and with
opposite
sign. Therefore
\begin{eqnarray}
Tr[{d\over{d\mu}}h(\mu) h^{-1}(s\mu)] &=&
{-1\over {{\pi}i}}\int dx_2 [\alpha_+(x_2)\partial_{2}\alpha_-(x_2)-
                 \alpha_-(x_2)\partial_{2}\alpha_+(x_2)]\nonumber\\
&&={1\over {2{\pi}^2}}\int dx_2 dy_2 \frac {(\alpha(y_2)-
\alpha(x_2))^2}{(x_2-y_2)^2}.
\label{formanab}
\end{eqnarray}
Eq.(\ref{formanab})coincides up to a factor of 2 with the
expresion
given by the Szego's Theorem for each one of the diagonal blocks. The
regularization factors for the determinant are $[\alpha]_{++}$ and
$[\alpha]_{--}$ and
we see that  they cancell with each other  in a natural way. This expression
coincides
with the result obtained by Falomir et al.
\cite{falomir},  using a perturbative approach. Note that our result does not
involve any
approximation.
By inserting this result in Eq.~(\ref{forman}) and recalling that
$\alpha=\theta+s_2$
we obtain
\begin{equation}
{d\over {d\mu}}ln\;\left (
\left.{{{\rm Det}( i \slp )_{B{\cal U}^{-1}(\mu +
\tau)}}
\over {{{\rm Det}( i \slp)}_{B{\cal U}^{-1}(\mu)}}}
\right)\right|_{\mu=0,\tau=1} =
{1\over {2{\pi}^2}}\int dx_2 dy_2 \frac
{[(\theta+s_2)(y_2)-(\theta+s_2)(x_2)]^2}
{(x_2-y_2)^2}.
\end{equation}
Following Falomir et al. \cite{falomir} we obtain the value of the ratio of the
determinants
which is given by
\begin{equation}
ln\;\left( {\frac{
{\rm Det} {( i \slp )}_{B{\cal U}^{-1}}}
{{\rm Det}( i \slp)_B}}\right)=
{1\over {4{\pi}^2}}\int dx_2 dy_2 \frac
{[(\theta+s_2)(y_2)-(\theta+s_2)(x_2)]^2}{(x_2-y_2)^2}.
\label{eq:finalforeman}
\end{equation}

This kind of non-local expressions have been used by Anderson\cite{Anderson}
and
Nozieres and Dominicis in the X-ray absorption
and emission in metals\cite{Nozieres2}and
 in the context of the Kondo problem\cite{Kondo}.
We have then computed the last ingredient we needed in Eq~(\ref{pf}) to
complete
the
bosonization scheme. In the next section we describe the bosonized theory
starting from
that expression for the partition function.

\section{Bosonized Theory}
\label{sec:currents}

Once the fermionic generating functional is known, any correlation function can
be
obtained by functional diferentiation. In particular, vacuum expectation values
of arbitrary products of fermion currents
\begin{equation}
<\prod_{i=1}^N \; J_{\mu}(x_i)>=<\prod_{i=1}^N \; {\bar \psi}(x_i) \gamma_\mu
\psi(x_i)>
\label{currents}
\end{equation}
can be obtained from the fermionic partition function
\begin{equation}
{\cal Z}[A]= \int {\cal D} {\bar \psi} {\cal D} \psi
\exp \left(-\int d^2x {\bar \psi}\;( i\slp +\not\!A)\psi\right)
\end{equation}
by using the regularization prescription for defining the fermion determinant,
as follows
\begin{equation}
<J_{\mu}(x)>=\left.{1\over{{\cal Z}[A]}}\frac{\delta {\cal Z}[A]}{\delta
A_\mu(x)}\right|_{A=0}.
\label{currentdef}
\end{equation}
Since, as we have shown in section II,  we have to split the external field
$A_{\mu}$
into two
fields, $B_\mu$ and $s_\mu$,
we define two currents, one in the bulk
\begin{equation}
<J_{\mu}(x)>=\left.{1\over{{\cal Z}[b_\mu,s_\mu]}}\frac{\delta {\cal
Z}[b_\mu,s_\mu]}{\delta
B_\mu(x)}\right|_{B,s=0}
\label{currentdefbulk}
\end{equation}
for $x_1>0$ and the other at the boundary
\begin{equation}
<J_{\mu}(0,x_2)>=\left.{1\over{{\cal Z}[b_\mu,s_\mu]}}\frac{\delta {\cal
Z}[b_\mu,s_\mu]}{\delta s_\mu(x_2)}\right|_{B,s=0}.
\label{currentdefbundary}
\end{equation}
By using Eq.~(\ref{pf}) and Eq.~(\ref{eq:finalforeman}), the partition function
is
\begin{eqnarray}
{\cal Z}[B_\mu,s_\mu]={\cal K} \int  {\cal D} \omega\;
&&\exp \left(-{1\over{2\pi}}\int (\partial_{\mu}\omega)^2
+ {i\over{\pi}}\int
{\epsilon}_{\mu\nu}\partial_{\nu}\omega(s_\mu+B_\mu)\right)\nonumber\\
&&\exp\left({1\over{-4\pi^2}} \int [\frac {(s_2(y_2)-s_2(x_2))^2}{(x_2-y_2)^2}-
\right.\nonumber\\
&&\;\;\;\;\left.\frac
{[(\theta+s_2)(y_2)-(\theta+s_2)(x_2)]^2}{(x_2-y_2)^2}\right)
\label{Pf}
\end{eqnarray}
which can be written as
\begin{eqnarray}
{\cal Z}[B_\mu,s_\mu]={\cal K} \int  {\cal D} \omega \;
&&\exp \left(-{1\over{2\pi}}\int (\partial_{\mu}\omega)^2 +{i\over{\pi}}\int
{\epsilon}_{\mu\nu}\partial_{\nu}\omega(s_\mu+B_\mu)\right)
\nonumber\\
&&\exp \left({1\over{2\pi}}\int dx_2dy_2\theta(x_2)K(x_2,y_2)\theta(y_2)\right)
\nonumber\\
&&\exp \left({1\over{\pi}}\int dx_2dy_2\theta(x_2)K(x_2,y_2)s_2(y_2)\right)
\label{pf1}
\end{eqnarray}
The kernel $K(x_2,y_2)$ is defined as
\begin{equation}
K(x_2,y_2)= \left( -{1\over{\pi}}{\cal P} \frac {1}{(x_2-y_2)^2}
+ {1\over{a}}\delta(x_2-y_2) \right).
\label{eq:kernel}
\end{equation}
Note that this action is already linear in the external fields. Then the
currents
can be
easly read from the partition function,
\begin{equation}
J_\mu(x)= {i\over{\pi}}\epsilon_{\mu\nu}\partial_{\nu}\omega\;\;,
\rm{for}\;x_1>0,
\label{Jbulk}
\end{equation}
which is the usual bosonization formula for the fermionic current in the bulk
(in
Euclidean space). The value of the current at the boundary is
\begin{equation}
J_\mu(0,x_2)={i\over{\pi}}\epsilon_{\mu\nu}\partial_{\nu}\omega|_{x_1=0} +
{\delta_{\mu2}\over{\pi}}\int dy_2\;K(x_2,y_2)\theta_{\mu}(y_2).
\label{Jborde}
\end{equation}
Let us now compute the current-current correlation functions.
They are given by
\begin{equation}
\left.<J_\mu(x)J_\nu(y)>\right|_{x_1,y_1>0}=\left.{1\over{{\cal
Z}[b_\mu,s_\mu]}}\frac{\delta^2
\ln{\cal Z}[b_\mu,s_\mu]}{\delta B_\mu(x_2)\delta B_\nu(y_2)}\right|_{B=0}
\end{equation}
\begin{equation}
\left.<J_\mu(x)J_\nu(y)>\right|_{x_1,y_1=0}=\left.{1\over{{\cal
Z}[b_\mu,s_\mu]}}\frac{\delta^2
\ln{\cal Z}[b_\mu,s_\mu]}{\delta s_\mu(x_2)\delta s_\nu(y_2)}\right|_{s=0}
\end{equation}
\begin{equation}
\left.<J_\mu(x)J_\nu(y)>\right|_{x_1>0,y_1=0}=\left.{1\over{{\cal
Z}[b_\mu,s_\mu]}}\frac{\delta^2
\ln{\cal Z}[b_\mu,s_\mu]}{\delta B_\mu(x_2)\delta s_\nu(y_2)}
\right|_{B,s=0}.
\end{equation}
Before computing the functional differentiation we have to integrate the
bosonic fields. Instead of using Eq.~(\ref{pf1}) for the partition function we
go a step
backwards and recall that the integral over $w$ comes from
Eq.~(\ref{eq:functionU2}) as
an alternative way of writing the Jacobian $J_F$ (Eq.~(\ref{omegaepsilon})). By
using
Eqs.~(\ref{eq:functint2}) and
(\ref{eq:finalforeman}) we
can also
express the partition function as
\begin{eqnarray}
{\cal Z}[B_\mu,s_\mu]=&&
\exp \left({1\over{2\pi}}\int (B_{\mu}(x) + s_{\mu}(x)) \Gamma_{\mu\nu}(x,y)
 (B_{\nu}(y)+
s_{\mu}(x))\right)\nonumber\\
&&\exp \left({1\over{2\pi}}\int dx_2dy_2\theta(x_2)K(x_2,y_2)\theta(y_2)\right)
\nonumber\\
&&\exp \left({1\over{\pi}}\int dx_2dy_2\theta(x_2)K(x_2,y_2)s_2(y_2)\right)
\end{eqnarray}

Then, the current-current correlation functions are
\begin{equation}
\left.<J_\mu(x)J_\nu(y)>\right|_{x_1,y_1>0}=
{1\over{\pi}} \Gamma_{\mu\nu}(x,y)
\end{equation}
\begin{equation}
\left.<J_\mu(x)J_\nu(y)>\right|_{x_1,y_1=0}=
{1\over {\pi}}\delta_{\mu,\nu}\delta_{\nu,2}K(x_2,y_2)
\end{equation}
\begin{equation}
\left.<J_\mu(x)J_\nu(y)>\right|_{x_1>0,y_1=0}=
{1\over{\pi}}{\Gamma}_{\mu\nu}(x,y)|_{x_1=0}.
\end{equation}
We follow now the method suggested in reference~\cite{laplata} to obtain
the
commutation relations from the current-current correlation functions at equal
time.

For  $x$, $y$ in the bulk, we find
\begin{eqnarray}
\left[J_2(x),J_1(y)\right]&=&\left.{1\over{2\pi}}(2{\Gamma}_{21}(x,y))\right|_{x_2=y_2}
\nonumber\\
&=&{1\over{\pi}}\partial_1\delta(x_1-y_1)
\end{eqnarray}
which is the usual commutation relation (up to a factor of $i$ which is
absent in Euclidean space).
Close to the boundary, the only non-vanishing commutator is
\begin{eqnarray}
\left[J_2(x),J_1(0,y_2)\right]&=&\left.{1\over{\pi}}(2{
\Gamma}_{21}(x,y)\right|_{x_2=y_2,y_1=0})\nonumber\\
&=&{1\over{\pi}}\partial_1\delta(x_1)
\end{eqnarray}
since at equal time $ \Gamma_{21}(x,y)|_{y_1=0}=-{1\over{2}}
\partial_1\delta(x_1)$.
Hence, we see that the presence of the boundary changes the currents but it
does
not change the commutation relations.

In next section we will see how the boundary and its degree of freedom
modify the one- particle Green's function.

\section{Bosonization of the Fermi operators}
\label{sec:bosofermiops}

	In this section we derive  a set of bosonization rules for the fermion
operators for systems with two types of boundary conditions: a) fixed
$R(0,x_2)=-L(0,x_2)$ and b) dynamical boundary conditions
$R(0,x_2)=-\exp(2\theta(x_2)) L(0,x_2)$, where $\theta(x_2)$ is a boundary
degree of
freedom. Our startegy is to first calculate
the one-particle Green's function in
 $1+{1\over{2}}$ dimensions with both types of boundary conditions and to use
these
results to determine the bosonization rules.
We will see that we can obtain a  generalization of the Mandelstam bosonization
rules for a semi-infinite system with specific boundary conditions.
In other words, we are going to find bosonic variables, $\omega(x)$ defined
in the
bulk, and $\theta(x_2)$ defined at the boundary, with an action
$S_B(\omega,\theta)$,
such that in the framework of this bosonic theory there exist operators with
the
same
 expectation values that those of the fermionic operators, $R$ and $L$ in the
framework of the fermionic theory.

\subsection{Boundary Condition R=-L}

The Fermionic one particle Green's function
is
\begin{equation}
S_{\alpha \beta}(x,y)=<\psi_\alpha(x)\bar{\psi}_\beta(y)>.
\end{equation}
and it satisfies  the Dirac equation
 \begin{equation}
 \left\{
\begin{array}{ll}
i\slp_x S_0(x,y)= \delta(x-y)\\
S_0^{11}(0,x_2;y)=-S_0^{21}(0,x_2;y)
\end{array}
\right.
\label {eq:S0}
\end{equation}
This boundary condition implies precisely $R(0,x_2)=-  L(0,x_2)$.
The solution for this differential equation is
\begin{equation}
S_0(x,y)=S_F(x,y)- \gamma_2S_F(x^{\star},y)
\end{equation}
where $S_F$ is the free fermionic Green's function, and $x^{\star}=(-x_1,x_2)$.
Explicitly,
\begin{equation}
S_0(x,y) \: ={1\over {2\pi}}
\: \left(
\begin{array}{cc}
-{1\over{w}} & -{1\over{\bar {z}}} \\
{1\over{z}} & {1\over{\bar {w}}}
\end{array}
\;\; \right)
\label{eq:S01}
\end{equation}
with $z=(x_1-y_1) +i(x_2-y_2)$ and $w=(-x_1-y_1) +i(x_2-y_2)$.

We will now proceed to find a bosonized expression for the fermion
operator consistent with the boundary condition $R=-L$.
The method that we follow is a generalization of the procedure used by
Fradkin and Kadanoff (FK) in
two-dimensional critical phenomena \cite{jefe,ceva}. It consists of
finding
 a set of operators in a theory
of a free bose field $\omega$ whose correlation function are given by
Eq.~(\ref{eq:S01}). Following FK, we define a bosonic
action coupled to a background gauge field as
\begin{equation}
S_B(\omega,A_\mu)= -{1\over {2\pi}}\int_\Omega d^2x \; (\partial_\mu
\omega + A_\mu)^2
\label{eq:SB}
\end{equation}
where $\omega$ is a bosonic variable defined in $\Omega$ which satisfies {\it
vanishing}
boundary conditions. Here $A_\mu$ is a vector field such that
\begin{equation}
B(z)\equiv {\epsilon}_{\mu\nu} \partial_\mu A_\nu(z)=\pi(-\delta(z-x) +
\delta(z-y)).
\label{eq:B}
\end{equation}
The gauge field $A_\mu$ represents two flux tubes, each
of flux $\pi$ and $-\pi$ respectively, at points $\vec x$ and $\vec y$.
In other terms, we have two {\it disorder operators}. Hereafter, we will
denote the insertion at $x$ of a disorder operator of flux (or
vorticity) $\pm \pi$ by the symbol ${\cal K}_{\pm \pi}(x)$. For {\it
infinite} systems, FK found that
the fermion operators  are represented  by suitable chosen linear
combinations
of products of disorder operators ${\cal K}_{\pm \pi}(x)$ and order (or
charge) operators $\exp(({\pm}i\omega(x'))$ (for $x \to x'$).
 This
construction is the euclidean analog of the Mandelstam
operators\cite{mandelstam} of abelian bosonization. We will now
investigate how does the presence of a boundary (as well as specific
boundary conditions) alter these rules.

We are interested in computing the expectation values of operators of the form
$e^{({\pm}i\omega(x){\pm}i\omega(y))}$ in the presence of disorder
operators ({\it i.e.\/} gauge fields of the from of Eq.~(\ref{eq:B})).
Therefore we need to compute bosonic partition functions of the form
\begin{equation}
{\cal Z}[A_\mu,J]= \int  {\cal D} \omega \; \exp\left(S_B + i\int_\Omega
d^2z \; J(z)\omega(z)\right)
\label{ZAJ}
\end{equation}
where the source $J$ has the form
\begin{equation}
J(z)=a\delta(z-x) + b\delta(z-y)
\label{J}
\end{equation}
and $a$,$b$ take the values $\pm 1$. We will show that the expectation values
of
these operators in the bosonic theory give the components of the one particle
fermionic Green's function (\ref{eq:S01}). Since the action above is quadratic
in
$\omega$ it can be integrated explicitly
\begin{eqnarray}
{\cal Z}[A_\mu,J]&=&{\cal K} \exp\left(-{1\over {2\pi}}\int_\Omega
d^2x \; A_\mu^2 \right)\nonumber\\
&&\exp\left(-{\pi\over {2}} \int_\Omega d^2x\; d^2y\; ({1\over
{\pi}}\partial_\mu A_\mu +iJ)(x)
G_0(x,y)
({1\over {\pi}}\partial_\mu A_\mu +iJ)(y)\right)
\end{eqnarray}
where the Green's function $G_0(x,y)$ was defined in Eq.~(\ref{eq:G0}). Note
that,
as it has been discused before, this Green's function is consistent with the
boundary
condition imposed on $\omega$.
After some algebra (notice that the integration by parts can be safely done
because
of the boundary conditions chosen) we get for ${\cal Z}[A_\mu,J]$
\begin{eqnarray}
{\cal Z}[A_\mu,J]&=&{\cal K}
\exp\left(-{1\over {2\pi}}\int_\Omega d^2x\; d^2y\; A_\mu(x) {\bar
{\Gamma}}_{\mu\nu}(x,y) A_\nu(y)\right)\nonumber\\
&&\exp\left({i\over {2}}\int_\Omega  d^2x\; d^2y\;
[A_\mu(x)\partial^{(x)}_\mu G_0(x,y) J(y) +
J(x)\partial^{(y)}_\nu G_0(x,y) A_\nu(y)]\right)\nonumber\\
&&\exp\left({\pi\over {2}}\int_\Omega d^2x\; d^2y\; J(x)G_0(x,y) J(y)\right)
\label{ZAJ2}
\end{eqnarray}
where the function $ {\bar {\Gamma}}_{\mu\nu}(x,y) $ is defined as
\begin{equation}
{\bar \Gamma}_{\lambda\alpha}(x,y)=
\left(\partial^{(x)}_{\mu}\partial^{(x)}_{\mu}\delta_{\alpha\lambda}+
\partial^{(x)}_{\lambda}\partial^{(y)}_{\alpha}\right)G_0(x,y)
\end{equation}
The Green's function of the half plane $G_0(x,y)$ is defined in terms of
$g_1$ and $g_2$, the real parts of the analytic
functions $\ln[(x_1-y_1) +i(x_2-y_2)]$ and $\ln[(-x_1-y_1) +i(x_2-y_2)]$
respectively.
The imaginary parts of these functions are
$\Phi_1(x,y)={1\over {2\pi}}\arctan(\frac {x_2-y_2}{x_1-y_1})$ and
$\Phi_2(x,y)={1\over {2\pi}}\arctan(\frac {x_2-y_2}{-x_1-y_1})$
respectively.
Therefore we can use the Cauchy-Riemann equations to get
\begin{equation}
\partial^{(x)}_\mu g_1(x,y)= \epsilon_{\mu\nu}\partial^{(x)}_\nu\Phi_1(x,y)
\label{Phi1}
\end{equation}
and
\begin{equation}
\partial^{(x)}_\mu g_2(x,y)=-\epsilon_{\mu\nu}
\partial^{(x)}_\nu\Phi_2(x,y).
\label{Phi2}
\end{equation}
With these relations we write the partition function as
\begin{eqnarray}
{\cal Z}[A_\mu,J]&=&{\cal K}
\exp\left(-{1\over {2\pi}}\int_\Omega  d^2x\; d^2y\; A_\mu(x) {\bar
{\Gamma}}_{\mu\nu}(x,y) A_\nu(y)\right)\nonumber\\
&&\exp\left({i\over {2}}\int_\Omega d^2x\; d^2y\;  [B(x) \Phi_1(x,y)J(y) +
J(x)\Phi_1(x,y)B(y) ]\right)\nonumber\\
&&\exp\left({i\over{2}}\int_\Omega d^2x\; d^2y\;  [ B(x) \Phi_2(x,y)J(y) -
J(x)\Phi_2(x,y)B(y) ]\right)\nonumber\\
&&\exp\left({\pi\over {2}}\int_\Omega d^2x\; d^2y\; J(x)G_0(x,y) J(y)\right)
\end{eqnarray}

We can use again the fact that $B(x)= \epsilon_{\mu\nu} \partial_\mu A_\nu(x)$
to obtain
\begin{eqnarray}
{\cal Z}[A_\mu,J]&=&{\cal K}
\exp\left({1\over {2}}\int_\Omega  d^2x\; d^2y\;  [{1\over {\pi}}B(x)
g_1(x,y)B(y) + \pi J(x)g_1(x,y)J(y)]\right)\nonumber\\
&&\exp\left({1\over {2}}\int_\Omega d^2x\; d^2y\;  [ i  B(x) \Phi_1(x,y)J(y) +
i J(x)\Phi_1(x,y)B(y)]\right)\nonumber\\
&&\exp\left({1\over {2}}\int_\Omega  d^2x\; d^2y\;  [{1\over {\pi}}B(x)
g_2(x,y)B(y)- \pi J(x)g_2(x,y)J(y)]\right)\nonumber\\
&&\exp\left({1\over {2}}\int_\Omega  d^2x\; d^2y\;  [ i  B(x)
\Phi_2(x,y)J(y) - i J(x)\Phi_2(x,y)B(y)]\right).
\label{ZAJ1}
\end{eqnarray}

Replacing  $J(z)= a\delta(z-x) + b\delta(z-y)$ and $B(z)=\pi( -\delta(z-x) +
\delta(z-y))$ in the integrands, the partition function
\begin{eqnarray}
{\cal Z}[A_\mu,J]&=&{\cal K}
\exp\left(\pi(-1+ab) g_1(x,y)+i\pi(-b+a)\Phi_1(x,y) - i\pi(b+a)
\Phi_2(x,y)\right)\nonumber\\
&&\exp\left({\pi\over {4}}(1-a^2)g_2(x,x) +{\pi\over {4}}(1-b^2)g_2(y,y) -
\pi(1+ab) g_2(x,y).
\right)
\end{eqnarray}

Then,
\begin{eqnarray}
\left.<\exp\left(ia \omega(x)+ib \omega(y)\right)>\right|_{bos}&&{\equiv}
\frac{{\cal Z}[A_\mu,J]}{{\cal Z}[0]}
\nonumber\\
&&=\exp\left(\pi(-1+ab) g_1(x,y)+i\pi(-b+a)\Phi_1( x,y)\right)\nonumber\\
&&\exp\left(-\pi(1+ab) g_2(x,y) -i\pi(b+a) \Phi_2(x,y)\right)\nonumber\\
&&\exp \left( {\pi \over 4}(1-a^2) g_2(x,x) +{\pi \over 4}(1-b^2)
g_2(y,y)\right)
\end{eqnarray}
There are four posible combinations of $a$ and $b$, which correspond to the
four matrix elements of $S_0(x,y)$. If $a=1,b=-1$ we get
\begin{eqnarray}
\langle\exp\left(i\omega(x)-i\omega(y)\right)\rangle&=&
\exp\left[-\ln \left((x_2-y_2)^2 +
(x_1-y_1)^2\right)^{1\over{2}}+\right.\nonumber\\
&&\left.+ i\arctan(\frac {x_2-y_2}{x_1-y_1}+ i\pi)\right]= -\frac{1}{\bar
z}.
\end{eqnarray}
Comparing with Eq.~(\ref{eq:S01}), we see that this term corresponds, in
the fermionic theory, to\hspace{\fill}
{$S^{12}_0(x,y)=<R(x)R^{\dagger}(y)>$}.
In the same way, with $a=b=-1$ we get
\begin{eqnarray}
\langle\exp\left(-i\omega(x)-i\omega(y)\right)\rangle&=&\exp\left[-\ln
\left((x_2-y_2)^2 +
(-x_1-y_1)^2\right)^{1\over{2}}+\right.\nonumber\\
&&\left. + i\arctan\left(\frac {x_2-y_2}{-x_1-y_1}\right)\right]= \frac{1}{\bar
w}
\end{eqnarray}
which corresponds to
$S^{22}_0(x,y)=<L(x)R^{\dagger}(y)>$.

Hence, we can identify
\begin{eqnarray}
R(x)&{\sim}&\;\;\exp(+i\omega(x))\;{\cal K}_{-\pi}(x) \nonumber \\
R^{\dagger}(x)&{\sim}&\;\;\exp(-i\omega(x))\;{\cal K}_{+\pi}(x)\nonumber \\
L(x)&{\sim}&\;\;\exp(-i\omega(x))\;{\cal K}_{+\pi}(x)\nonumber \\
L^{\dagger}(x)&{\sim}&\;\;\exp(+i\omega(x))\;{\cal K}_{-\pi}(x)
\label{eq:identRL}
\end{eqnarray}
where $\sim$ means
that we identify products of operators (in the sense of the operator
product expansion) inside arbitrary matrix elements of the fermionic and
bosonic theories respectively. In the same way we can identify the
remaining matrix elements.
We conclude that, for the case of the boundary condition $R=-L$, the
fermion operator on the half-plane is constructed in the same way as in
FK\cite{jefe}. The boundary condition appears only through the presence
of diagonal operators which mix $L$ and $R$ components of the Fermi
field.

It is worthwhile to note that, under global chiral (euclidean)
transformations, $R \to  e^{\theta_0}\; R$ and $L \to  e^{-\theta_0}\; L$, the
off-diagonal matrix elements of the fermion propagator remain invariant
but the diagonal matrix elements do not. In the bosonized theory,
global chiral transformations are represented by $\omega \to \omega
+\theta_0$. This representation of the matrix elements clearly satisfies
this symmetry. Notice, however, that a global chiral  transformation
requires that the boundary condition be modified since the boundary condition
itself
breaks the chiral symmetry explicitly. Below we will come back on this issue.

\subsection{Dynamical Boundary Conditions}

Up to this point, we have shown how, following FK, we can calculate the
one particle Green's
function
in its bosonized form  for the boundary conditions $R(x_2)=-L(x_2)$.
The next step is the calculation of the Green's function with dynamical
boundary conditions. In other words, we are looking for the solutions of
\begin{equation}
 \left\{
\begin{array}{ll}
i\slp_x S_F(x,y|\theta)= \delta(x-y)\\
S^{11}_F(0,x_2;y|\theta)=- e^{2\theta(x_2)}S^{21}_F(0,x_2;y|\theta) .
\end{array}
\right.
\label {eq:Stheta}
\end{equation}
Here, $\theta(x_2)$ represents a quantum mechanical degree of freedom
localized at the boundary, {\it i.~e.\/} a quantum impurity. For the
rest of this section we will take $\theta(x_2)$ to be arbitrary but
fixed. In the next section we will consider a model with a fully
dynamical, quantum mechanical boundary degree of freedom.

 The solution of Eq.~(\ref{eq:Stheta}) is
\begin{equation}
S_F(x,y|\theta) \:  ={\frac{1}{2\pi}}
 \left(
\begin{array}{cc}
-\frac{b(x)a^{-1}(y)}{w}
& -\frac{b(x)b^{-1}(y)}{\bar {z}}\\
\frac{a(x)a^{-1}(y)}{z}
& \frac{a(x)b^{-1}(y)}{\bar {w}}
\end{array}
\;\; \right)
\label{eq:poto}
\end{equation}
The functions $a(x)$ and $b(x)$ have to be chosen in such a way that the
differential equation and the boundary condition of
Eq.~(\ref{eq:Stheta}) are satisfied. In particular,
it has to be ($\partial_{x_2}+ i\partial_{x_1})b(x)=
(\partial_{x_2}- i\partial_{x_1})a(x)=0$.

To find the explicit form of $a(x)$ and $b(x)$ we will use the FK approach.
In the method
developed in the previous subsection, the bosonic field  satisfies vanishing
boundary
conditions. This reflects the fact that the boundary conditions for the
fermions
were $R(0,x_2) =- L(0,x_2)$. But we have yet to determine how dynamical
boundary conditions may affect
the boundary conditions for the bosons.
Our strategy will be to relate $S_F(x,y|\theta)$
with a Green's function that satisfies non-twisted, $R=-L$,
 boundary conditions. For that
purpose we define $S(x,y)$ as a solution of the following equation
\begin{equation}
 \left\{
\begin{array}{ll}
(i\slp_x + \slbarA)S(x,y)= \delta(x-y)\\
S^{11}(0,x_2;y)=-S^{21}(0,x_2;y)
\end{array}
\right. .
\label{eq:S}
\end{equation}
Now, it is obvious that we can find $S(x,y)$ by using the method described
in the previous subsection.
To  establish a relation between $S(x,y)$ and $S_F(x,y|\theta)$ we write
${\bar {A_{\mu}}}$ as
\begin{equation}
{\bar {A_{\mu}}}= \epsilon_{\mu\nu}\partial_{\nu}\rho
\end{equation}
where $\rho(x)$ is a function such that $\rho(0,x_2)=\theta(x_2)$.
It can be seen that if we take,
\begin{equation}
S_F(x,y|\theta)= e^{\gamma_5\rho(x)}S(x,y)e^{\gamma_5\rho(y)}.
\label{SStheta}
\end{equation}
then $S_F(x,y|\theta)$ is a solution of Eq.(\ref{eq:Stheta}).
Following the same steps that lead to the calculation of $S_0$, we define
\begin{eqnarray}
{\cal Z}[A_\mu,J,{\bar {A_\mu}}]=&& \int  {\cal D} \omega \;
\exp S_B(\omega,A_\mu) \nonumber\\
&&\exp\left( i\int_\Omega J(z)\omega(z) + \int_\Omega J_\mu(z){\bar
{A_\mu}}(z)\right)\nonumber\\
&&\exp \left({1\over{2\pi}}\int
dx_2dy_2\theta(x_2)K(x_2,y_2)\theta(y_2)+\right.\nonumber\\
&&\left.{1\over{\pi}}\int dx_2dy_2{\bar A}_2(x_2)K(x_2,y_2)\theta(y_2)\right)
\end{eqnarray}
Here, $A_{\mu}$ and $J$ are the external sources defined in Eqs.~(\ref{eq:B})
and
{}~(\ref{J}) respectively, and $J_\mu(z)$ is the current defined in
(\ref{Jbulk}). If we define $S_B(\theta,{\bar A}_2)$ as
\begin{equation}
S_B(\theta,{\bar A}_2)={1\over{2\pi}}\int
dx_2dy_2\theta(x_2)K(x_2,y_2)\theta(y_2) +
{1\over{\pi}}\int dx_2dy_2{\bar A}_2(x_2)K(x_2,y_2)\theta(y_2)
\end{equation}
 the partition function can be written as
\begin{eqnarray}
{\cal Z}[A_\mu,J,{\bar {A_\mu}}]&&= e^{S_B(\theta,{\bar A}_2)}\;\int  {\cal D}
\omega
\;\exp\left(S_B(\omega,A_\mu)+
 i\int_\Omega [J(z) -{1\over{\pi}}\epsilon_{\mu\nu}\partial_\nu
{\bar {A_\mu}}(z)]\omega(z)\right)\nonumber\\
&&=e^{S_B(\theta,{\bar A}_2)}\;\int  {\cal D} \omega \;\exp\left(S_B +
 i\int_\Omega {\tilde{J}}(z)\omega(z)\right)
\end{eqnarray}
with $ {\tilde{J}}(z)=J(z) -{1\over{\pi}}\epsilon_{\mu\nu}
\partial_\nu{\bar {A_\mu}}(z)$. This expression  is formally equal to
Eq.~(\ref{ZAJ}).
 Hence, we can follow the same steps that took  this equation into
Eq.~(\ref{ZAJ2})
to obtain
\begin{eqnarray}
{\cal Z}[A_\mu,J,{\bar {A_\mu}}]&&=e^{S_B(\theta,{\bar A}_2)}\; {\cal
Z}[A_\mu,J]\nonumber\\
&&\exp\left({i\over{2\pi}}\int_\Omega[{\partial_\mu}{A_\mu}(x)G_0(x,y)
{\nabla}^2\rho(y)+
{\nabla}^2\rho(x) G_0(x,y){\partial_\mu}{A_\mu}(y)]\right)\nonumber\\
&&\exp\left({-1\over{2}}\int_\Omega[J(x)
G_0(x,y){\nabla}^2\rho(y)+{\nabla}^2\rho(x)
 G_0(x,y) J(y)]\right)\nonumber\\
&&\exp\left({1\over{2\pi}}\int {\bar A}_{\mu}(x)
\Gamma_{\mu\nu}(x,y){\bar A}_{\nu}(y)\right)
\end{eqnarray}
where ${\cal Z}[A_\mu,J]$ is exactly equal to  Eq.~(\ref{ZAJ1}) and the other
 factors come from the extra term in ${\tilde{J}}(x)$. These two factors can be
computed with the same techniques used before giving
\begin{eqnarray}
{\cal Z}[A_\mu,J,{\bar {A_\mu}}]&&=e^{S_B(\theta,{\bar A}_2)
+{1\over{2\pi}}\int {\bar
A}_{\mu}(x) \Gamma_{\mu\nu}(x,y){\bar A}_{\nu}(y)}
{\cal Z}[A_\mu,J]\nonumber\\
&&\exp\left(-a\rho(x) -b\rho(y)\right)\;
\exp\left({1\over{\pi}}\int\;dz_2\; \rho(0,z_2) \;\frac {ax_1 +
i(x_2-z_2)}{x_1^2 +
(x_2-z_2)^2}
\right)\nonumber\\
&&\exp\left({1\over{\pi}}\int\;dz_2\; \rho(0,z_2) \;\frac {by_1 -
i(y_2-z_2)}{y_1^2 +
 (y_2-z_2)^2}\right).
\end{eqnarray}
Then,
\begin{eqnarray}
\langle\exp\left(ia \omega(x)+ib\omega(y)\right)\rangle&&{\equiv}\frac{{\cal Z}
[A_\mu,J,{\bar {A_\mu}}]}{{\cal Z}[{\bar {A_\mu}}]}
\nonumber\\
&&=\exp\left(\pi(-1+ab) g_1(x,y)+i\pi(-b+a)\Phi_1( x,y)\right)\nonumber\\
&&\exp\left(-\pi(1+ab) g_2(x,y) -i\pi(b+a) \Phi_2(x,y)\right)\nonumber\\
&&\exp\left({1\over{\pi}}\int\;dz_2\; \rho(0,z_2)\; \frac {ax_1 +
i(x_2-z_2)}{x_1^2 + (x_2-z_2)^2}\right)\nonumber\\
&&\exp\left({1\over{\pi}}\int\;dz_2\; \rho(0,z_2) \;\frac {by_1 -
 i(y_2-z_2)}{y_1^2 + (y_2-z_2)^2}\right)\nonumber\\
&&\exp\left(-a\rho(x) -b\rho(y)\right)
\end{eqnarray}
In this case also, there are four combinations of $a$ and $b$ which give the
four matrix elements of $S(x,y)$. Up to a normalization constant the one
particle
Green's function is
\begin{equation}
S(x,y) \: =\: {\frac{1}{2\pi}}
\left(
\begin{array}{cc}
-\frac{e^{-\rho(x)}b(x)a^{-1}(y)e^{-\rho(y)}}{w}
& -\frac{e^{-\rho(x)}b(x)b^{-1}(y)e^{\rho(y)}}{\bar {z}}
\\
\frac{e^{\rho(x)}a(x)a^{-1}(y)e^{-\rho(y)}}{z}
& \frac{e^{\rho(x)}a(x)b^{-1}(y)e^{\rho(y)}}{\bar
{w}}
\end{array} \;\; \right)\;
\end{equation}
where
\begin{equation}
a(x)= \exp\left({1\over{\pi}}\int\;dz_2\; \theta(z_2)\;\frac {-x_1 +
 i(x_2-z_2)}{x_1^2 + (x_2-z_2)^2}\right)
\label{eq:a}
\end{equation}
\begin{equation}
b(y)=\exp\left({1\over{\pi}}\int\;dz_2\; \theta(z_2)\; \frac {y_1 +
i(y_2-z_2)}{y_1^2 + (y_2-z_2)^2}\right).
\label{eq:b}
\end{equation}
and
\begin{equation}
\rho(x)={1\over{\pi}}\int\;dz_2\; \theta(z_2) \frac {x_1}{x_1^2 +
(x_2-z_2)^2}.
\label{eq:rhodef}
\end{equation}

Note that $\rho(x_1,x_2)$ satisfies
\begin{equation}
 \left\{
\begin{array}{ll}
\nabla^2 \rho(x)=0 \;\;\;$\rm in$\;\; x_1>0\\
\rho(0,x_2)=\theta(x_2)
\end{array}
\right.
\end{equation}

It can be seen that
\begin{eqnarray}
\ln a(x)&=& -2\int_{\omega<0} {d\omega \over {2\pi}}\rho(\omega)
e^{i\omega(x_2-ix_1)} \nonumber\\ &&\equiv -2\rho_-(x_2-ix_1)
\end{eqnarray}
Hence, $\ln a(x)$ is the anti-analytic extension of the negative
frequency part of $\rho(0,x_2)=\theta(x_2)$.
In the same way,
\begin{eqnarray}
\ln b(x)&=& 2\int_{\omega>0} {d\omega \over {2\pi}}\rho(\omega)
e^{i\omega(x_2+ix_1)} \nonumber\\ &&\equiv 2\rho_+(x_2+ix_1)
\end{eqnarray}
is the analytic extension of the positive frequency part of
$\rho(0,x_2)=\theta(x_2)$.
Note that $\rho(x_1,x_2)$ is the real part of both, $\rho_+(x_1+ix_2)$
and $\rho_-(x_1- ix_2)$ and that
\begin{equation}
2\rho_+(x_2+ix_1) +2\rho_-(x_2-ix_1)=2\rho(x_1,x_2)
\label{suma}
\end{equation}

Using the relation between
$S_F(x,y|\theta)$ and $S(x,y)$ (Eq.~(\ref{SStheta})), we get an expression in
the
bosonized theory  for the fermion Green's function which satisfies dynamical
boundary
conditions (Eq.~(\ref{eq:poto})).

By using Eqs.(\ref{eq:a}-\ref{eq:b}) we can rewrite  Eq.(\ref{eq:poto}) as
\begin{equation}
S_F(x,y|\theta) \:  ={\frac{1}{2\pi}}
 \left(
\begin{array}{cc}
-\frac{e^{{1\over {\pi}}\int dz_2 \theta(z_2)h_{11}(z_2)}}{w}
& -\frac{e^{{1\over {\pi}}\int dz_2 \theta(z_2)h_{12}(z_2)}}{\bar {z}}\\
\frac{e^{{1\over {\pi}}\int dz_2 \theta(z_2)h_{21}(z_2)}}{z}
& \frac{e^{{1\over {\pi}}\int dz_2 \theta(z_2)h_{22}(z_2)}}{\bar {w}}
\end{array}
\;\; \right)
\label{eq:potoint}
\end{equation}
where
\begin{eqnarray}
h_{11}(z_2)&=&\frac{1}{x_1-i(x_2-z_2)}+\frac{1}{y_1+i(y_2-z_2)}\nonumber\\
h_{12}(z_2)&=&\frac{1}{x_1-i(x_2-z_2)}-\frac{1}{y_1-i(y_2-z_2)}\nonumber\\
h_{21}(z_2)&=&-\frac{1}{x_1+i(x_2-z_2)}+\frac{1}{y_1+i(y_2-z_2)}\nonumber\\
h_{22}(z_2)&=&-\frac{1}{x_1+i(x_2-z_2)}-\frac{1}{y_1-i(y_2-z_2)}
\label{eq:haches}
\end{eqnarray}

We can summarize these results in the form of a set of bosonization
rules for the
fermion operators, which now include the effects of the presence of a
boundary degree of freedom.
As in the previous case, {\it i.~e.\/} for $R(x_2)=-L(x_2)$, we can identify
the
fermionic operators,
 $R(x_2)$ and $L(x_2)$, with the corresponding bosonic operators. From
Eq.~(\ref{eq:poto})
 we see that
\begin{eqnarray}
R(x_2)&{\sim}&\;\;\exp\left(i\omega(x) + {1\over{\pi}}\int\;dz_2\; \theta(z_2)
\frac {x_1
+i(x_2-z_2)}{x_1^2 + (x_2-z_2)^2}\right)\;\; {\cal K}_{-\pi}(x)\nonumber \\
R^{\dagger}(x_2)&{\sim}&\;\;\exp\left(-i\omega(x) - {1\over{\pi}}\int\;dz_2\;
\theta(z_2)
\frac {x_1+i(x_2-z_2)}{x_1^2 + (x_2-z_2)^2}\right)\;\; {\cal
K}_{+\pi}(x)\nonumber \\
L(x_2)&{\sim}&\;\;\exp\left(-i\omega(x) + {1\over{\pi}}\int\;dz_2\; \theta(z_2)
\frac {-x_1+
 i(x_2-z_2)}{x_1^2 + (x_2-z_2)^2}\right)\;\;{\cal K}_{+\pi}(x)\nonumber \\
L(x_2)&{\sim}&\;\;\exp\left(+i\omega(x) - {1\over{\pi}}\int\;dz_2\; \theta(z_2)
\frac {-x_1+ i(x_2-z_2)}{x_1^2 + (x_2-z_2)^2}\right)\;\;{\cal K}_{-\pi}(x)
\label{eq:idtwist}
\end{eqnarray}
where, as before, $\sim$ means that the bosonic operators yield the same
correlation
functions (with the same boundary conditions) as the operators $R$ and $L$
in the Fermi theory. This result completes the bosonization construction.
Notice that
the main modification is the presence of a {\it boundary operator} in the
definition
of the fermion operator of this generalization of Mandelstam's construction.
We should stress here that, as remarked in ~\ref{sec:Fermions},
in order to compare with the operator construction in
Minkowski space, not only the imaginary time $x_2$ needs to be Wick-rotated
back to
the real time $x_0$ but the boundary degree of freedom $\theta$  also has to be
continued to the imaginary axis, $\theta \to i \theta$.

\section{Free fermions coupled to a dynamical boundary degree of freedom}
\label{sec:dynamical}

Now we will compute the one particle Green's function in the
case that
$\theta$ is a quantum mechanical degree of freedom. That is, instead of
regarding
$\theta(x_2)$ as a prescribed classical parameter, it has to be regarded of as
an extra degree of freedom of the full system. In order to account for the
effects
of the dynamics of $\theta$, we have to specify its dynamics through an extra
term
in the lagrangian and we must also integrate over
all  possible configurations of $\theta$. Hence, the full one particle Green's
function
becomes
\begin{eqnarray}
S_F(x,y)&=&\frac {\int {\cal D} {\bar \psi} {\cal D} \psi {\cal D} \theta
\;\psi(x){\bar
\psi}(y)\;e^{iS({\bar \psi}, \psi,\theta) +i
S_\theta(\theta)}} {\int {\cal D} {\bar \psi} {\cal D} \psi {\cal D} \theta
e^{iS({\bar \psi},
\psi,\theta) +
iS_\theta(\theta)}}\nonumber\\
&&=\frac {\int  {\cal D} \theta\; S_F(x,y|\theta)\; {\rm Det}(\theta)\;e^{i
S_\theta(\theta)}}
{\int {\cal D} \theta \;{\rm Det}(\theta)\;e^{iS_\theta(\theta)}}
\label{eq:dynamicalgf}
\end{eqnarray}
where $S({\bar \psi}, \psi,\theta)$ is the action for the masless fermions in
the half line coupled to a dynamical degree of freedom at the boundary.
${\rm Det}(\theta)$ is the determinant calculated in Section III
(Eq.~(\ref{eq:finalforeman})). Upon exponentiation, it yields an induced term
in the
action $ S_{\rm ind}(\theta)$ of the form
\begin{equation}
S_{\rm ind}(\theta)\equiv
ln\;\left({\frac{{\rm Det} (i \slp )_{B{\cal U}^{-1}}}
{{\rm Det}( i \slp )_B}}\right)=
{1\over {4{\pi}^2}}\int dx_2 dy_2 \frac
{[\theta(y_2)-\theta(x_2)]^2}{(x_2-y_2)^2+a^2}.
\label{eq:S-ind-theta}
\end{equation}
which represents the dynamics of $\theta$ {\it induced}
by the fermionic degree of freedom. The intrinsic dynamics of the boundary
degree of
freedom is represented by the term $S_\theta(\theta)$ in the action.

At this point, it is convenient (and helpful to understand the physics) to make
an
analytic continuation of the boundary degree of freedom $\theta \to i \theta$.
In this
way, we recover the usual periodicity property $\theta \to \theta+2 \pi$ . In
Euclidean space this symmetry is not manifest since the group of chiral
transformations is non-compact in Euclidean space. This change has for effect
to  (a)
change the sign of the induced action $S_{ind}(\theta)$ and (b) the Green
function
amplitudes need to be analytically continued (Eq.(~\ref{eq:potoint})).

The intrinsic dynamics of the boundary degree of freedom $\theta$ is determined
by the
underlying microscopic system from which the Luttinger-type model is derived. A
simple and still realistic model of this dynamics consists in regarding
$\theta$ as a
coordinate for a single degree of freedom moving on a ring. This is the
situation in
the Caldeira-Leggett model~\cite{caldeira} which describes the effects of a
local
degree of feedom coupled to a ``fermion bath". From the point of view of a
Hubbard
model or a quantum wire, $\theta$ represents a boundary effect of
electron-electron correlations. A set of ``interesting" terms that can be
incorporated
in $S_\theta(\theta)$ are (in imaginary time)
\begin{equation}
S_\theta(\theta)=
\int dx_2 \left( {\frac{M}{2}}\left({\frac{\partial \theta}{\partial
x_2}}\right)^2
-V( \theta) \right)
\label{eq:Sthetaprop}
\end{equation}
where $M$ is the mass of the boundary degree of freedom.
In quantum wire problems,  the potential  $V \left( \theta \right)$ is the
backscattering amplitude at the boundary.  This term is
of the form
\begin{equation}
V( \theta)= G \cos ( \beta \theta )
\label{eq:Vtheta}
\end{equation}
where $\beta$ is a parameter which, if interactions
are present, depends on the electron-electron coupling constant.

By simple inspection we can see that all terms of the action, except for the
potential
term, are invariant under a global chiral rotation of the boundary conditions,
$\theta
(x_2) \to \theta(x_2)+\theta_0$. The potential term reduces this continuous
symmetry
to a discrete subgroup $\theta(x_2) \to \theta(x_2)+2\pi n/\beta$ , where $n$
is an
arbitrary integer. Similarly, the {\it off-diagonal} terms of the Green
function are
{\it invariant} under global chiral rotations whereas the {\it diagonal} terms
{\it are not}. These symmetries have a very simple interpretation in the
bosonized theory. Since the chiral symmetry of the fermions translates into the
invariance of the bosonized theory by constant shifts of the  field $\omega$,
the
boundary condition  $\omega=0$ at $x_1=0$ reflects the fermion boundary
condition
$R=-L$. The boundary degree of freedom $\theta$ can be viewed as the value of
$\omega$
at the boundary. Hence, a path integral in which one integrates over all values
of
$\theta$ without restrictions should correspond to free boundary conditions for
$\omega$. We will see below that this is indeed the case and that, in the
absence of
potential terms, the boundary conditions are free and the system behaves as in
the
case of an infinite line. Conversely, potential terms act to ``pin" the
boundary
value of the bosonic field $\omega$  to a finite set of possible values. This
behavior
is closely related to the picture of Affleck and Ludwig of the Kondo problem as
a
crossover in boundary conditions~\cite{Affleck}.

In the framework of macroscopic quantum tunneling and coherence,
Fisher and Zwerger~\cite{fisherzwerger} have studied systems with actions of
this same
type. By means of renormalization group arguments, they concluded that the
kinetic
energy term is {\it irrelevant} at low frequencies (or long times), that the
induced
term in the action is a strictly marginal operator and that the cosine terms of
the
potential are relevant. Fisher and Zwerger refer to this strong-coupling fixed
point as
to the {\it localization transition}. The induced term in the action plays the
same role
in this problem as the free massless scalar theory (or Gaussian model) does in
two-dimensional critical systems and in conformal field theory.

We are interested in the computation of the matrix elements of the fermion
one-particle Green's function for a system with dynamical boundary conditions.
To
simplify the discussion we will include the effects of the potential term only
within
a linearized approximation in which  $V( \theta)\approx - {\frac{G
\beta^2}{2}}
 \theta^2 \equiv -{\frac{\bar G}{2}} \theta^2 $. This approximation misses the
important periodicity property and overemphasizes the physics of localization.
Recent
work by Ludwig and colaborators~\cite{ludwig} and by Tsvelik~\cite{Tsvelik}
indicates
that there is interesting physics which this fixed point does not describe
correctly.

Hence, we conclude that the effective dynamics of the boundary degree of
freedom
$\theta$ is described by the action
\begin{equation}
S_{\rm eff}(\theta)=
{1\over {2\pi}} \int dx_2dy_2 \theta(x_2) K(x_2,y_2) \theta(y_2)+
\int dx_2 \left( {\frac{M}{2}}\left({\frac{\partial \theta}{\partial
x_2}}\right)^2
+{\frac{\bar G}{2}} \theta^2 \right)
\label{eq:Sefftheta}
\end{equation}
where $K(x_2,y_2)$ is given by Eq.(\ref{eq:kernel}).
In order to compute the one-particle fermion Green's function we need to
evaluate functional integrals over $\theta$ of the form
\begin{equation}
W_{\alpha \beta}\equiv
\langle \exp \left( {i\over {\pi}} \int dz_2 \theta(z_2)
h_{\alpha \beta}(z_2)\right) \rangle_\theta ={\frac{
\int {\cal D} \theta \; \exp\left( -S_{\rm eff}(\theta)
+ {i\over {\pi}} \int dz_2 \theta(x_2) h_{\alpha \beta}(z_2)\right)}
{\int {\cal D} \theta \; \exp\left( -S_{\rm eff}(\theta)\right)}}
\label{eq:vev}
\end{equation}
Since the exponent is quadratic in $\theta$, these integrals can be done
explicitly.
In Fourier space, the action $S_{\rm eff}(\theta)$ is
\begin{equation}
S_{\rm eff}(\theta)=\int {\frac{d\omega}{2\pi}} |\theta(\omega)|^2
{\frac{K_{\rm eff}(\omega)}{2\pi}}
\end{equation}
where the kernel $K_{\rm eff}(\omega)$ is given by
\begin{equation}
K_{\rm eff}(\omega)={\frac{1-e^{-a|\omega|}}{a}}+\pi M \omega^2+\pi
{\bar G}
\label{eq:Keff}
\end{equation}
The form of the kernel $K_{\rm eff}(\omega)$ shows that the inertial term acts
like a high frequency cutoff.  At low frequencies we can use the approximation
$K_{\rm eff}(\omega)\approx |\omega|+\pi {\bar G}$.
The expectation values of the form of Eq.(~\ref{eq:vev}) become
\begin{equation}
W_{\alpha \beta}= \exp(-{\frac{1}{2\pi}}\int
{\frac{d\omega}{2\pi}}h_{\alpha \beta}(-\omega)
{\frac{1}{K_{\rm eff}(\omega)}} h_{\alpha \beta}(\omega))
\label{eq:path}
\end{equation}
where $h_{\alpha \beta}(\omega)$ is the Fourier transform  of
$h_{\alpha \beta}(z_2)$.
By making use of the  expressions for the amplitudes
$h_{\alpha \beta}(z_2)$ given in Eq.(~\ref{eq:haches}), we find that their
Fourier transforms can be written in terms of the functions
\begin{equation}
F_{\pm}(\omega;u,v)=\int_{-\infty}^{+\infty} d\tau
{\frac{e^{i \omega \tau}}{u \pm i( \tau-v)}}
= 2\pi e^{-|\omega| (u \mp i v))} \Theta( \pm \omega)
\end{equation}
where $u >0$ and $\Theta(x)$ is the step function.
The Fourier transforms become
\begin{eqnarray}
h_{11}(\omega)&=&F_+(\omega;x_1,x_2)+F_-(\omega;y_1,y_2)=
2\pi e^{-|\omega|(x_1-ix_2)} \Theta(\omega)+
2\pi e^{-|\omega|(y_1+iy_2)} \Theta(-\omega)
\nonumber \\
h_{12}(\omega)&=&F_+(\omega;x_1,x_2)-F_+(\omega;y_1,y_2)=
2\pi \left(e^{-|\omega|(x_1-ix_2)}-e^{-|\omega|(y_1-iy_2)}\right)
\Theta(\omega)
\nonumber \\
h_{21}(\omega)&=&-F_-(\omega;x_1,x_2)+F_-(\omega;y_1,y_2)=
2\pi \left(-e^{-|\omega|(x_1+ix_2)}+e^{-|\omega|(y_1+iy_2)}\right)
\Theta(-\omega)
\nonumber \\
h_{22}(\omega)&=&-F_-(\omega;x_1,x_2)-F_+(\omega;y_1,y_2)=
-2\pi e^{-|\omega|(x_1+ix_2)} \Theta(-\omega)-
2\pi e^{-|\omega|(y_1-iy_2)} \Theta(\omega)
\label{eq:h}
\end{eqnarray}
We can now write down an explicit
formula for the logarithm of the amplitudes $W_{\alpha \beta}$. We find
that the {\it off-diagonal} matrix elements vanish identically, $\ln
W_{12}=\ln W_{21}\equiv 0$ and that the {\it diagonal} matrix elements
are given by
\begin{equation}
\ln {\bar W}= \ln W_{11}=\ln W_{22}=-2\int_0^\infty d\omega
{\frac{e^{-\omega( (x_1+y_1)-i(x_2-y_2))}}{K_{\rm eff}(\omega)}}
\label{eq:lnW11}
\end{equation}
The explicit computation of the integral in eq.~\ref{eq:lnW11} yields the
result
\begin{equation}
\ln {\bar W}=2 e^{\pi {\bar G} {\bar w}} Ei(-\pi {\bar G} {\bar w})
\end{equation}
where $Ei(-x)$ is the exponential-integral function. This result
yields the asymptotic behaviors of $\ln {\bar W}$ as
\begin{eqnarray}
\ln {\bar W}& \approx& 2 \left( {\bf C} +
\ln (-{\bar w}\pi {\bar G})\right),\;\; \mbox{for $|w|\pi {\bar G} \ll
1$} \nonumber \\
\ln {\bar W} &\approx& -{\frac{2}{\pi {\bar G}{\bar w}}}
, \;\; \mbox{for $|w|\pi {\bar G} \gg 1$}
\label{eq:wbar}
\end{eqnarray}
where ${\bf C}$ is Euler's constant.
In the limit $x_1+y_1 \to 0$, these results hold provided that $x_1+y_1$ is
replaced by
the smallest of the cutoff $a$ and $\sqrt{\pi M}$.

With these results at hand we find that, for dynamical boundary
conditions,
the matrix elements of the Fermion Green's function are given by
\begin{equation}
S_F(x,y) \:  = {\frac{1}{2\pi}}
 \left(
\begin{array}{cc}
-{\frac{{\bar W}}{w}}
& -{\frac{1}{{\bar z}}}\\
{\frac{1}{z}}
& {\frac{{\bar W}}{\bar w}}
\end{array}
\;\; \right)
\label{eq:potofinal}
\end{equation}
where ${\bar {W}}$ is given in Eq.(~\ref{eq:lnW11}).

Hence, we find that the off
diagonal matrix elements of the fermion Green's function, {\it i.~e.\/} the
propagators  $\langle R(x) R^\dagger(y)\rangle$ and  $\langle L(x)
L^\dagger(y)\rangle$
respectively,  are {\it equal} to their value in the infinite plane and
are unaffected
by dynamical boundary conditions. Notice, however, that for a {\it specified}
configuration of the boundary chiral angle $\theta(x_2)$, these matrix elements
are
modified.
In contrast, we find that the diagonal matrix elements,
{\it i.~e.\/} the
propagators $\langle R(x) L^\dagger(y)\rangle$ and   $\langle L(x)
R^\dagger(y)\rangle$
respectively, are modified by dynamical boundary conditions.  However,
these changes
are only significant either close to the boundary and at short  times.
At long times
and far away from the boundary, the off diagonal matrix elements
behave exactly as
in the case of $R=-L$ boundary conditions. Close to the boundary,  we
find power law corrections of the form
\begin{equation}
\langle R(x) L^\dagger(y)\rangle \sim -{\frac{{\rm const.}(\pi {\bar
G}{\bar w})^2}{2\pi w}}
\end{equation}
If the potential term is absent (${\bar G} \to 0$), the boundary degree
of  freedom $\theta$ is unpinned. In this regime, the
{\it diagonal} matrix elements of the fermion one-particle  Green's
function vanish identically, even away from the boundary,
\begin{equation}
S_F(x,y) \:  = {\frac{1}{2\pi}}
\: \left(
\begin{array}{cc}
0
& -\frac{1}{\bar {z}}\\
\frac{1}{z}
& 0
\end{array}
\;\; \right)\;
\end{equation}
In this
limit the boundary  degree of freedom fluctuates wildly and it
effectively removes the
boundary condition $\omega=0$ for the bosonized theory. In other
terms, the bosonized
theory has {\it free} boundary conditions instead of Dirichlet. In this regime,
the
fermions are not scattered by the boundary.
In a sense, they are ``eaten" by the
boundary  degree of freedom. The fermion Green's function in this limit
is equal to the
Green's function in the infinite plane ({\it i.~e.\/} the full line).
Once again, this result agrees with the picture of Affleck and Ludwig.

Finally, let us note that the amplitude ${\bar W}$ has a clear  physical
meaning.
If ${\bar W}$ was a constant, it would have the straightforward physical
interpretation as a {\it backscattering amplitude}. However, it is both
position and time-dependent. Hence, the backscattering
amplitude determined from ${\bar W}$ is both momentum and energy (frequency)
dependent. The actual form of this dependence is rather involved, as can
be seen by inspecting the form of ${\bar W}$ in real time. The frequency
dependence of
$\bar W$
is simply a manifestation of the crossover between the two asymptotic behaviors
shown in eq.~\ref{eq:wbar}. Indeed, at frequencies high compared with $1/(\pi
{\bar
G})$, $\bar W$ scales to zero at low frequancies, {\it i.~e.\/} it  behaves
asymptotically as in the case in which the boundary  degree of freedom is
unpinned
(${\bar G}) \to 0$). In this regime the effective value of the diagonal matrix
elements of the fermion Green's function scale to zero and the effective
boundary
condition is free.  Conversely, at low frequencies and for $\bar G$ fixed,
$\bar W$
scales to a constant. This is the behavior of the pinned boundary  degree of
freedom
and it amounts to a boundary condition $R=-L$ for the fermions.

\section{Conclusions}
\label{sec:conc}

In this paper we have reexamined the bosonization of a theory of free fermions
in the
presence of boundaries and with specific boundary conditions. This is a very
simple
system in which the role of boundary conditions and of boundary degrees of
freedom can
be investigated quite explicitly.

We have exhibited the role of the boundary conditions of the fermions  in the
bosonized theory. We used the Forman's method for the computation of
fermion determinants to derive the action of the bosonized theory which
exhibits the
effects of the boundary degrees of freedom. This action allowed us to
determine
the form of the currents and found that the currents acquire additive
corrections due to the boundary degree of freedom. These corrections are
needed in order to to satisfy the conservation laws.
In contrast, we find that the current algebra is not affected by the presence
of
boundaries and of boundary degrees of freedom. We also constructed the
bosonized form
of the Fermi operators . We showed that the boundary degrees of freedom enter
explicitly in the definition of the Fermi operators. Finally, we used these
results to
calculate the fermion one-particle Green's function for a theory with a
dynamical
boundary degree of freedom.

The methods described in this paper can be generalized so as to include
the effects of interactions and of non-abelian symmetries.
It is also possible to use these methods to investigate the
interesting non-perturbative structure recently found , even in abelian
systems,
by means of the thermodynamic Bethe Ansatz~\cite{ludwig,Tsvelik}.
In a separate publication~\cite{enviasdedesarrollo} we apply our methods
to the case of interacting fermions, in particular to the Luttinger model and
study
the interplay between the Kondo effect and interactions in $1+1$-dimensional
systems.

\section{Acknowledgements}

This work was supported in part
by the National Science Foundation through the grants NSF DMR-91-22385 at
the University of Illinois at Urbana-Champaign, NSF DMR-89-20538
at the Materials Research Laboratory of the University of Illinois(MF,EF), by a
Glasstone Research Fellowship in Sciences and a Wolfson Junior Research
Fellowship (AL) and by the Ministerio de Educaci\`on y Ciencia, Spain(EM).

\newpage
\appendix
\section{Boundary Jacobian}
\label{sec:AA}

The aim of this appendix is the computation of
\begin{eqnarray}
\ln{\delta}J_{F}&=&\lim_{s, \epsilon \rightarrow0}\left( \int_0^1 dt\;\int
dx_2\;\int^\epsilon_0
dx_1< x | e^{-s \slD_t \slD_t} | x>
\phi(x_1,x_2) \right)\nonumber\\
&&=-2\lim_{s, \epsilon \rightarrow0}\left( \int_0^1 dt\;\int
dx_2\;\int^\epsilon_0 dx_1tr[K_2(i
\slD_t;x,x)\gamma_5\phi(x_1,x_2)] \right).
\end{eqnarray}
In order to do this computation we have to find the heat kernel $K_2$ in
$(0,\epsilon)$.
The operator $i\slD$ can be written as
\begin{eqnarray}
i\slD\: &=&
\: \left(
\begin{array}{cc}
0 & -\partial_1 + B \\
\partial_1 + B & 0
\end{array} \; \right)\;
\end{eqnarray}
with $B=i\partial_2 + A_2$.
Following APS\cite{APS} we make the aproximation
$A_2(x_1,x_2){\approx}A(0,x_2)$. In
order to compute the trace, we need to find a basis of eingestates of $i\slD_t
$. We
construct such a basis as follows. Let us call $e_\lambda(x_2)$ the eigenvector
of
eigenvalue $\lambda$ of the operator $B$. That is,
\begin{equation}
Be_\lambda(x_2)={\lambda}e_\lambda(x_2).
\end{equation}
It can be shown that there exist $w_{k,\lambda}$ and a set of functions
$f_{k,\lambda}(x_1)$ and $g_{k,\lambda}(x_1)$ such that
\begin{equation}
\left\{
\begin{array}{ll}
i\slD  \psi_{k,\lambda}(x_1,x_2)=w_{k,\lambda} \psi_{k,\lambda}(x_1,x_2)\\
f_{k,\lambda}(0)=-g_{k,\lambda}(0)
\end{array}
\right.
\label{slD}
\end{equation}
where by definition  $\psi_{k,\lambda}(x_1,x_2)$
is
\begin{eqnarray}
 \psi_{k,\lambda}(x_1,x_2)\: &=&
\: \left(
\begin{array}{cc}
f_{k,\lambda}(x_1)e_\lambda(x_2)\\
g_{k,\lambda}(x_1)e_\lambda(x_2)
\end{array} \;\; \right)\nonumber\\
&&\equiv
\: \left(
\begin{array}{cc}
R_{k,\lambda}(x_1,x_2)\\
L_{k,\lambda}(x_1,x_2)
\end{array} \;\; \right).
\end{eqnarray}
Specifically,
\begin{equation}
\psi_{k,\lambda}(x_1,x_2) \: =\frac{e_\lambda(x_2)}{(2(\lambda +
w_{k,\lambda})w_{k,\lambda})^{1\over 2}}
\: \left(
\begin{array}{cc}
(\lambda + w_{k,\lambda})\sin kx_1 - k \cos kx_1\\
(\lambda + w_{k,\lambda})\sin kx_1 + k \cos kx_1
\end{array}
\;\; \right)
\end{equation}
form the complete set that satisfies Eq.~(\ref{slD}) if we take $w_{k,\lambda}=
(k^2 +
\lambda^2)^{1\over 2}$ ($k\geq 0$), and $\lambda= {2\pi \over {T}}( n + {1\over
2} +
\Lambda)$
where $\Lambda$ is the flux of the gauge field through the boundary.

Then,
\begin{eqnarray}
\ln{\delta}J_{F}=\lim_{s, \epsilon \rightarrow0}\left( -2 \int
dx_2\;\phi(0,x_2)\;\int^\epsilon_0
dx_1\right.&&\left.[\sum_{k,\lambda} R_{k,\lambda}^{\dagger}(x_1)\phi
e^{-s(\partial_1^2 +
B^2)}R_{k,\lambda}(x_1) -\right.\nonumber\\
&&\left.\sum_{k,\lambda} L_{k,\lambda}^{\dagger}(x_1)\phi e^{-s(\partial_1^2 +
B^2)}L_{k,\lambda}(x_1)]\right)\nonumber\\
=\lim_{s, \epsilon \rightarrow0}\int dx_2 \phi(0,x_2) \int_0^\infty
\frac{dk}{\pi}
\sum_\lambda &&\left(\frac{e^{-s(k^2 + \lambda^2)}}{(k^2 + \lambda^2)^{1\over
2}}\right)
\cos(2k\epsilon -1).
\end{eqnarray}
It is important to note that in the above limit, $s \ll \epsilon $.  In the
limit $T\rightarrow
\infty$ the sum over $\lambda$ can be replaced by an integral. By using the
Euler's
formula it can be shown that
\begin{equation}
\ln{\delta}J_{F}={-2T\over {\pi^2}}\lim_{s, \epsilon \rightarrow0}\int dx_2\;
\phi(0,x_2)\int_0^\infty \frac{dk}{\pi} \int_0^\infty du \left(\frac{e^{-s(k^2
+
u^2)}}{(k^2 +
u^2)^{1\over 2}}\right) \cos(2k\epsilon -1)
\end{equation}
The result of this last integral is
\begin{eqnarray}
\ln{\delta}J_{F}={-T\over {4\pi}}\lim_{s, \epsilon \rightarrow0}\int dx_2\;
\phi(0,x_2){\Gamma({1\over 2})}{\sqrt s}&&\left\{ e^{({\delta^2 \over {4}}+
\epsilon^2){1\over
s}}\left(1- \Phi(\sqrt{({\delta^2 \over {4}}+ \epsilon^2)}{1\over {\sqrt
s}})\right) -
\right.\nonumber\\
&&\left.-e^{({\delta^2 \over {4s}})}\left(1- \Phi({\delta^2 \over  {2\sqrt
s}})\right)\right\}
\label{J_F}
\end{eqnarray}
where $\delta$ is a Schwinger parameter that goes to zero and $\Phi(x)$
is the error
function. The order of the limits is
$\delta \ll s \ll \epsilon$. Then in the first term of Eq.~(\ref{J_F}) we use
the asymptotic
expansion of $\Phi(\sqrt x )$. But in the second term we use the limit for
$x\ll
0$ of
$\Phi(x)$. Hence Eq.~(\ref{J_F}) becomes
\begin{equation}
\ln{\delta}J_{F}={-T\over {4\pi}}\lim_{s, \epsilon \rightarrow0}\int dx_2
\;\phi(0,x_2)\frac{\Gamma({1\over 2})}{\sqrt s}\left( \frac{\Gamma({1\over
2}){\sqrt
s}}{\epsilon} -1\right).
\label{J_F2}
\end{equation}
The second term in Eq.~(\ref{J_F2}) tells us that there is a singularity in
$s$.
According to
the definition of the heat kernel regularization\cite{laplata} we have to keep
only the finite
part of (\ref{J_F2}). Therefore, the properly regularized jacobian at the
boundary is
\begin{equation}
{\delta}J_{F} =\exp\left( -{T\over{4\epsilon}}
\int dx_2 \;\phi(0,x_2)\right).
\end{equation}
Since $\phi(0,x_2) =0$, the contribution of this jacobian to the partition
function is
one.
\section{Forman's Determinant}
\label{sec:BB}

In this appendix we are going to set up the notation and and show some of the
calculations that lead to eq.\ (\ref{traza})
First note that we are working in the space spanned by $e^{ iw_{n} x_0}$ with
$n$ runing over the integers. This space splits naturally into the one
expanded by $e^{ iw_{n} x_0}$ with $n\geq0$ (the space of functions with
positive frequencies) and the space spanned by $e^{ iw_{n} x_0}$ with
$n<0$.
 Indeed, each $\alpha$ can be decomposed into $\alpha_{+}$ and $\alpha_{-}$
such that $\alpha = \alpha_{+}+\alpha_{-}$ and
\begin{eqnarray}
\alpha_{+}(x_2)&=& \sum_{n\geq 0} \alpha_{n} e^{ iw_{n} x_2}\nonumber\\
\alpha_{-}(x_2)&=& \sum_{n<0}\alpha_{n} e^{ iw_{n} x_2}.
\end{eqnarray}
In the same spirit, we can decompose operators $\cal A$ acting in such space
into blocks:
\begin{equation}
{\cal A} \: =
\: \left(
\begin{array}{cc}
{[{\cal A}]}_{--} & {[{\cal A}]}_{-+} \\
{[{\cal A}]}_{+-} & {[{\cal A}]}_{++}
\end{array} \;\; \right)\;
\end{equation}
where ${[{\cal A}]}_{--}$ means ${[{\cal A}]}_{m-n}$ with $m,n <0$ and so on.

Finally, we use the notation $\alpha\equiv \alpha(\mu,x_2)\equiv
\mu\alpha(x_2)$
The first step is the calculation of the
inverse of
\begin{equation}
h(\mu) \: =
\: \left(
\begin{array}{cc}
{[e^{-\alpha}]}_{--} & {[e^{\alpha}]}_{-+} \\
{[e^{-\alpha}]}_{+-} & {[e^{\alpha}]}_{++}
\end{array} \;\; \right)\;.
\end{equation}

Formally it has the form
\begin{equation}
h^{-1}(\mu) \: =
\: \left(
\begin{array}{cc}
{[U]}_{--} & {[U]}_{-+} \\
{[U]}_{+-} & {[U]}_{++}
\end{array} \;\; \right)\;
\end{equation}

where
\begin{eqnarray}
{[U]}_{--} &=& {\{1-
               {({[e^{-\alpha}]}_{--})}^{-1}{[e^{\alpha}]}_{-+}
               {({[e^{\alpha}]}_{++})}^{-1}{[e^{-\alpha}]}_{+-}\}}^{-1}
               {({[e^{-\alpha}]}_{--})}^{-1}\;,\nonumber \\
{[U]}_{++} &=& {\{1-
               {({[e^{\alpha}]}_{++})}^{-1}{[e^{-\alpha}]}_{+-}
               {({[e^{-\alpha}]}_{--})}^{-1}{[e^{\alpha}]}_{-+}\}}^{-1}
               {({[e^{\alpha}]}_{++})}^{-1} \;,\nonumber \\
{[U]}_{-+} &=& {{({[e^{-\alpha}]}_{--})}^{-1}{[e^{\alpha}]}_{-+}
               \{{({[e^{\alpha}]}_{++})}^{-1}{[e^{-\alpha}]}_{+-}
               {({[e^{-\alpha}]}_{--})}^{-1}{[e^{\alpha}]}_{-+}
               -1\}}^{-1}{({[e^{\alpha}]}_{++})}^{-1}\;,\nonumber \\
{[U]}_{+-} &=& {{({[e^{\alpha}]}_{++})}^{-1}{[e^{-\alpha}]}_{+-}
               \{{({[e^{-\alpha}]}_{--})}^{-1}{[e^{\alpha}]}_{-+}
               {({[e^{\alpha}]}_{++})}^{-1}{[e^{-\alpha}]}_{+-}
               -1\}}^{-1}{({[e^{-\alpha}]}_{--})}^{-1}\;.
\end{eqnarray}

We need to compute
\begin{eqnarray}
Tr[{d\over{d\mu}}h(\mu) h^{-1}(\mu)]&&
=Tr\left( {d\over {d\mu}}({[e^{-\alpha}]}_{--}){[U]}_{--} +
    {d\over {d\mu}}({[e^{\alpha}]}_{-+}){[U]}_{+-}\right.\nonumber\\
&+&\left.
    {d\over {d\mu}}({[e^{-\alpha}]}_{+-}){[U]}_{-+}
    +{d\over {d\mu}}({[e^{\alpha}]}_{++}){[U]}_{++}\right)= \nonumber\\
&&Tr\;\left\{\left({d\over {d\mu}}({[e^{-\alpha}]}_{--})-{d\over {d\mu}}
({[e^{\alpha}]}_{-+}){({[e^{\alpha}]}_{++})}^{-1}
{[e^{-\alpha}]}_{+-}\right)\right.\nonumber\\
     &&\left.{\{1 -{({[e^{-\alpha}]}_{--})}^{-1}
{[e^{\alpha}]}_{-+}{({[e^{\alpha}]}_{++})}^{-1}
      {({[e^{-\alpha}]}_{+-})}\}}^{-1}{({[e^{-\alpha}]}_{--})}^{-1}\right\}
\nonumber\\
    &+& Tr\;\left\{\left({d\over {d\mu}}({[e^{\alpha}]}_{++})-
{d\over {d\mu}}
({[e^{-\alpha}]}_{+-}){({[e^{-\alpha}]}_{--})}^{-1}{[e^{\alpha}]}_{-+}
\right)\right.\nonumber\\
      &&\left.{\{1 -{({[e^{\alpha}]}_{++})}^{-1}{[e^{-\alpha}]}_{+-}
{({[e^{-\alpha}]}_{--})}^{-1}
{({[e^{\alpha}]}_{-+})}\}}^{-1}{({[e^{\alpha}]}_{++})}^{-1}\right\}.
\end{eqnarray}
In order to compute this last expression it is usefull to derive some
identities
relating the matrix elements used above. Note that for deriving this
expressions
we are going to use the fact that $\alpha$ is an abelian variable. Appart from
that $\alpha$ could be any antiperiodic function in $[0,\beta]$.
The main propertie that we are using here is the fact that
${[e^{\alpha_{-}}]}_{p}=0$ for $p\geq0$ and
${[e^{\alpha_{+}}]}_{n}=0$ for $n<0$. Having this in mind the following
identities are straightforward
\begin{eqnarray}
{[e^{\alpha_{-}}]}_{+-}&=&0\\
{[e^{\alpha_{+}}]}_{-+}&=&0\\
{[e^{\alpha_{-}}]}_{++}{[e^{\alpha_{+}}]}_{++} &=&
                                [e^{\alpha_{-}}e^{\alpha_{+}}]_{++}\\
{[e^{\alpha_{-}}]}_{++}{[e^{-\alpha_{-}}]}_{++} &=& 1_{++}\\
{[e^{\alpha_{+}}]}_{++}{[e^{-\alpha_{+}}]}_{++} &=& 1_{++}\\
{[e^{\alpha_{+}}]}_{--}{[e^{\alpha_{-}}]}_{--} &=&
                                 [e^{\alpha_{+}}e^{\alpha_{-}}]_{--}\\
{[e^{\alpha_{+}}]}_{--}{[e^{-\alpha_{+}}]}_{--} &=& 1_{--}\\
{[e^{\alpha_{-}}]}_{--}{[e^{-\alpha_{-}}]}_{--} &=& 1_{--}\\
{[e^{\alpha_{-}}]}_{-+}{[e^{\alpha_{+}}]}_{++} &=&
                 [e^{\alpha_{-}}e^{\alpha_{+}}]_{-+}\\
{[e^{\alpha_{+}}]}_{--}{[e^{\alpha_{-}}]}_{-+}&=&
                 [e^{\alpha_{+}}e^{\alpha_{-}}]_{-+}\\
{[e^{\alpha_{-}}]}_{++}{[e^{\alpha_{+}}]}_{+-}&=&
                 [e^{\alpha_{-}}e^{\alpha_{+}}]_{+-}\\
{[e^{\alpha_{+}}]}_{+-}{[e^{\alpha_{-}}]}_{--}&=&
                 [e^{\alpha_{+}}e^{\alpha_{-}}]_{+-}\\
{[e^{\alpha_{+}}]}_{++}{[e^{-\alpha_{+}}]}_{+-}&=&
            -{[e^{\alpha_{+}}]}_{+-}{[e^{-\alpha_{+}}]}_{--}\\
{[e^{\alpha_{-}}]}_{--}{[e^{-\alpha_{-}}]}_{-+}&=&
            -{[e^{\alpha_{-}}]}_{-+}{[e^{-\alpha_{-}}]}_{++}.
\label{ffm}
\end{eqnarray}
With a change $\alpha \rightarrow -\alpha$ more identities can be derived.
It is inmediate to realize that
\begin{eqnarray}
{({[e^{-\alpha}]}_{--})}^{-1}&=&{[e^{\alpha_{-}}]}_{--}
{[e^{\alpha_{+}}]}_{--}\nonumber\\
{({[e^{\alpha}]}_{++})}^{-1}&=&{[e^{-\alpha_{+}}]}_{++}
{[e^{-\alpha_{-}}]}_{++}.
\end{eqnarray}
As an example of how to use the identities (B7)-(B20) we compute the following
expression
$ {\{1
-{({[e^{-\alpha}]}_{--})}^{-1}{[e^{\alpha}]}_{-+}{({[e^{\alpha}]}_{++})}^{-1}
      {({[e^{-\alpha}]}_{+-})}\}}^{-1}$.

We define ${[M]}_{--}$ as
\begin{equation}
{[M]}_{--}\equiv
{({[e^{-\alpha}]}_{--})}^{-1}{[e^{\alpha}]}_{-+}{({[e^{\alpha}]}_{++})}^{-1}
      {[e^{-\alpha}]}_{+-}.
\end{equation}
By using Eq.(B21) we obtain
\begin{equation}
{[M]}_{--}=
{[e^{{\alpha}_-}]}_{--}{[e^{{\alpha}_+}]}_{--}{[e^{\alpha}]}_{-+}
{[e^{{-\alpha}_+}]}_{++}{[e^{{-\alpha}_-}]}_{++}{[e^{-\alpha}]}_{+-}
\end{equation}
With Eqs.(B15) and (B18) we can write the factors of the form
$[e^{\alpha}]$ in terms of $[e^{{\alpha}_-}]$ and $[e^{{\alpha}_+}]$
\begin{eqnarray}
{[M]}_{--}&=&
{[e^{{\alpha}_-}]}_{--}{[e^{{\alpha}_+}]}_{--}{[e^{{\alpha}_-}]}_{-+}
{[e^{{\alpha}_+}]}_{++}
{[e^{{-\alpha}_+}]}_{++}{[e^{{-\alpha}_-}]}_{++}{[e^{{-\alpha}_+}]}_{+-}
{[e^{{-\alpha}_-}]}_{--}=\nonumber\\
&=&{[e^{{\alpha}_-}]}_{--}{[e^{{\alpha}_+}]}_{--}{[e^{{\alpha}_-}]}_{-+}
{[e^{{-\alpha}_-}]}_{++}{[e^{{-\alpha}_+}]}_{+-}
{[e^{{-\alpha}_-}]}_{--}.\nonumber
\end{eqnarray}

After rearranging some factors using Eqs.(B16) and (B17) ${[M]}_{--}$
can be written as follows
\begin{eqnarray}
{[M]}_{--}=
{[e^{{\alpha}_-}]}_{--}{[e^{\alpha}]}_{-+}
{[e^{-\alpha}]}_{+-}{[e^{{-\alpha}_-}]}_{--}.\nonumber
\end{eqnarray}

Hence
\begin{eqnarray}
(1-{[M]}_{--})&=&
{[e^{{\alpha}_-}]}_{--}(1-{[e^{\alpha}]}_{-+}{[e^{-\alpha}]}_{+-})
{[e^{{-\alpha}_-}]}_{--} \nonumber\\
&=&{[e^{{\alpha}_-}]}_{--}{[e^{\alpha}]}_{--}{[e^{-\alpha}]}_{--}
{[e^{{-\alpha}_-}]}_{--}.
\end{eqnarray}
By using eqs.(B14) the inverse of the above expression reads
\begin{eqnarray}
{(1-{[M]}_{--})}^{-1}={[e^{{-\alpha}_-}]}_{--}
{({[e^{-\alpha}]}_{--})}^{-1}{({[e^{\alpha}]}_{--})}^{-1}
{[e^{{\alpha}_-}]}_{--}.
\end{eqnarray}
The factors ${({[e^{-\alpha}]}_{--})}^{-1}$ and ${({[e^{\alpha}]}_{--})}^{-1}$
can be computed with the help of eqs.(B12)-(B14). The result is
\begin{eqnarray}
{({[e^{-\alpha}]}_{--})}^{-1}{({[e^{\alpha}]}_{--})}^{-1}=
{[e^{{\alpha}_-}]}_{--}{[e^{{\alpha}_+}]}_{--}{[e^{{-\alpha}_-}]}_{--}
{[e^{{-\alpha}_+}]}_{--}.
\end{eqnarray}
Therefore, the expression we were calculating turns out to be
\begin{eqnarray}
\{1-{({[e^{-\alpha}]}_{--})}^{-1}{[e^{\alpha}]}_{-+}&&{({[e^{\alpha}]}_{++})}^{-
1}
      {[e^{-\alpha}]}_{+-}\}=\nonumber\\
&&={[e^{\alpha_{-}}]}_{--}{[e^{\alpha_{-}}]}_{--}{[e^{\alpha_{+}}]}_{--}
{[e^{-\alpha_{-}}]}_{--}{[e^{-\alpha_{+}}]}_{--}
{[e^{-\alpha_{-}}]}_{--}.
\end{eqnarray}

Following the same steps we can see that
\begin{eqnarray}
\{1-{({[e^{\alpha}]}_{++})}^{-1}{[e^{-\alpha}]}_{+-}&&{({[e^{-\alpha}]}_{--})}
^{-1}{[e^{\alpha}]}_{-+}\}=\nonumber\\
&&={[e^{-\alpha_{+}}]}_{++}{[e^{-\alpha_{+}}]}_{++}{[e^{-\alpha_{-}}]}_{++}
{[e^{\alpha_{+}}]}_{++}{[e^{\alpha_{-}}]}_{++}{[e^{\alpha_{+}}]}_{++}.
\end{eqnarray}

The next step is the calculation of the derivatives.
\begin{eqnarray}
({d\over {d\mu}}{[e^{\alpha}]}_{n-m})&=&
-{\int} e^{-2\pi i (n-m)x_2}\alpha(x_2)e^{-\mu\alpha(x_2)} \nonumber\\
 &=&-\sum_{p} {[e^{-\mu\alpha(x_0)}]}_{n-m-p}[\alpha]_{p}\nonumber\\
 &=&-\sum_{q} {[e^{-\mu\alpha(x_0)}]}_{n-q}[\alpha]_{q-m}.
\end{eqnarray}
 And in our notation,
\begin{equation}
({d\over {d\mu}}{[e^{\alpha}]}_{--})=-({[e^{-\alpha}]}_{--}[\alpha]_{--}
                           + {[e^{-\alpha}]}_{-+}[\alpha]_{+-}).
\end{equation}
 In the same way,
\begin{eqnarray}
{d\over {d\mu}}{[e^{-\alpha}]}_{-+}&=& ({[e^{\alpha}]}_{--}[\alpha]_{-+}
                           + {[e^{\alpha}]}_{-+}[\alpha]_{++})\nonumber\\
{d\over {d\mu}}{[e^{\alpha}]}_{+-}&=& -({[e^{-\alpha}]}_{+-}[\alpha]_{--}
                           + {[e^{-\alpha}]}_{++}[\alpha]_{+-})\nonumber\\
{d\over {d\mu}}{[e^{-\alpha}]}_{++}&=& ({[e^{\alpha}]}_{+-}[\alpha]_{-+}
                           + {[e^{\alpha}]}_{++}[\alpha]_{++}).
\end{eqnarray}

By inserting (B27)-(B31) into (B6) we obtain
\begin{eqnarray}
&&Tr[{d\over{d\mu}}h(\mu) h^{-1}(\mu)]
=\nonumber\\
&&Tr\;[\alpha]_{-+}\{
{[e^{-\alpha_{+}}]}_{++}{[e^{-\alpha_{+}}]}_{++}{[e^{-\alpha_{-}}]}_{++}
{[e^{\alpha_{+}}]}_{++}{[e^{\alpha}]}_{+-}-\nonumber\\
&&\;\;\;\;\;\;\;\;\;{[e^{-\alpha_{+}}]}_{++}{[e^{-\alpha_{-}}]}_{++}
{[e^{-\alpha_{+}}]}_{+-}
{[e^{\alpha_{-}}]}_{--}{[e^{\alpha_{+}}]}_{--}{[e^{-\alpha_{-}}]}_{--}
{[e^{\alpha}]}_{--}\}\;+\nonumber\\
&&Tr\;[\alpha]_{++}\{
{[e^{-\alpha_{+}}]}_{++}{[e^{-\alpha_{+}}]}_{++}{[e^{-\alpha_{-}}]}_{++}
{[e^{\alpha_{+}}]}_{++}{[e^{\alpha}]}_{++} - \nonumber\\
&&\;\;\;\;\;\;\;\;\;{[e^{-\alpha_{+}}]}_{++}{[e^{-\alpha_{-}}]}_{++}
{[e^{-\alpha_{+}}]}_{+-}
{[e^{\alpha_{-}}]}_{--}{[e^{\alpha_{+}}]}_{--}{[e^{-\alpha_{-}}]}_{--}
{[e^{\alpha}]}_{-+}\}\;+\nonumber\\
&&Tr\;[\alpha]_{+-}\{
{[e^{\alpha_{-}}]}_{--}{[e^{\alpha_{+}}]}_{--}{[e^{\alpha_{-}}]}_{-+}
{[e^{-\alpha_{+}}]}_{++}{[e^{-\alpha_{-}}]}_{++}{[e^{\alpha_{+}}]}_{++}
{[e^{-\alpha}]}_{++}-\nonumber\\
&&\;\;\;\;\;\;\;\;\;{[e^{\alpha_{-}}]}_{--}{[e^{\alpha_{-}}]}_{--}
{[e^{\alpha_{+}}]}_{--}
{[e^{-\alpha_{-}}]}_{--}{[e^{-\alpha}]}_{-+}\}\;\nonumber\\
&&Tr\;[\alpha]_{--}\{
{[e^{\alpha_{-}}]}_{--}{[e^{\alpha_{+}}]}_{--}{[e^{\alpha_{-}}]}_{-+}
{[e^{-\alpha_{+}}]}_{++}{[e^{-\alpha_{-}}]}_{++}{[e^{\alpha_{+}}]}_{++}
{[e^{-\alpha}]}_{+-}-\nonumber\\
&&\;\;\;\;\;\;\;\;\;{[e^{\alpha_{-}}]}_{--}{[e^{\alpha_{+}}]}_{--}
{[e^{\alpha_{+}}]}_{--}
{[e^{-\alpha_{-}}]}_{++}{[e^{-\alpha}]}_{--}\}.\nonumber\\
\end{eqnarray}
By using the identities (B7)-(B20) we can see that the factor of
$[\alpha]_{--}$ contributes with $-1$ and the factor of $[\alpha]_{++}$
contibutes with 1 . The other two
terms are more complicated to calculate but it can be done. The final result
is
\begin{eqnarray}
Tr[{d\over{d\mu}}h(\mu) h^{-1}(\mu)]&=&
tr \left\{ [\alpha]_{++} -  [\alpha]_{--} \right.\nonumber\\
&+& [\alpha]_{+-}( -{[e^{{\alpha}_-}]}_{--} {[e^{{-\alpha}_{-}}]}_{-+} +
{[e^{{\alpha}_{-}}]}_{--}{[e^{{\alpha}_{-}}]}_{-+}{[e^{{-\alpha}_{-}}]}_{++}
{[e^{{-\alpha}_{-}}]}_{++} ) \nonumber\\
&+&\left.
[\alpha]_{-+}({[e^{{-\alpha}_+}]}_{++}{[e^{{\alpha}_+}]}_{+-} -
{[e^{{-\alpha}_{+}}]}_{++}{[e^{{-\alpha}_{+}}]}_{+-} {[e^{{\alpha}_{+}}]}_{--}
{[e^{{\alpha}_{+}}]}_{--}) \right\}.
\end{eqnarray}

 \section{Euclidean-Minkowski Correspondences for Dirac Fermions}
\label{sec:CC}

Upon the analytic continuation from real time $x_0$ to  imaginary time
$x_2=i x_0$, a
number of quantities are transformed. Firstly, the Minkowski
space metric $g_{\mu \nu}$
\begin{equation}
g_{\mu \nu}=
\left(
\begin{array}{cc}
1  & 0  \\
0 & -1
\end{array}
\right)
\label{eq:minkmetric}
\end{equation}
becomes, in euclidean space
\begin{equation}
g_{\mu \nu}=-\delta_{\mu \nu}.
\label{eq:euclmetric}
\end{equation}
The Minkowski space $\gamma$-matrices are (in the  chiral basis)  are
\begin{equation}
 \gamma_0= \sigma_1\qquad \gamma_1 =-i \sigma_2  \qquad  \gamma_5=
\sigma_3
\label{eq:minkowskigamma}
\end{equation}
and have the property
\begin{equation}
{\rm tr} (\gamma_\mu \gamma_\nu)= 2\;g_{\mu \nu}.
\label{eq:trace}
\end{equation}
The euclidean $\gamma$-matrices are chosen to be hermitean, in which
case they are
\begin{equation}
\gamma_1 =\sigma_2 \qquad \gamma_2= \sigma_1  \qquad  \gamma_5=
\sigma_3
\label{eq:euclideangamma}
\end{equation}
and they satisfy the relations
\begin{equation}
{\rm tr} (\gamma_\mu \gamma_\nu)= 2\;\delta_{\mu \nu}, \qquad
\gamma_5 \gamma_\mu=-i \epsilon_{\mu \nu} \gamma_\nu.
\label{eq:relations}
\end{equation}

We also give the correspondence for
\begin{equation}
{\bar \psi}=\psi^{\dagger}\; \gamma_0 \to i {\bar \psi}
\label{eq:psibar}
\end{equation}
as well as for vector fields, which transform like
\begin{equation}
A_0 \to i A_2 \qquad A_1 \to A_1.
\label{eq:vector}
\end{equation}
With this notation the covariant derivatives become
\begin{eqnarray}
i \slp+\lnA =
\left(
\begin{array}{cc}
0  & \partial_{\bar z}-iA_{\bar z}  \\
-\partial_{z}+iA_{\bar z} & 0
\end{array}
\right)
\label{eq:dslash}
\end{eqnarray}
where we have set
\begin{eqnarray}
\begin{array}{lcr}
\partial_{ z}= \partial_1-i \partial_2 &,&
\partial_{\bar z}=\partial_1+i \partial_2\\
A_{ z}=A_1-i A_2&,&
A_{\bar z}=A_1+i A_2 .
\end{array}
\label{eq:complex1}
\end{eqnarray}
Using these identifications the Minkowski space actions  ${\cal S}_M$
and ${\cal S}_E$ are related in the usual manner
\begin{equation}
i{\cal S}_M \to -{\cal S}_E.
\label{eq:actionmapping}
\end{equation}
For the specific case of
a massive Dirac fermion coupled to a background gauge field, the  Lagrangian
${\cal L}_{M}$,
\begin{equation}
{\cal L}_{M}=
{\bar \psi} \; (i  \slp+\lnA ) \; \psi- M {\bar \psi} \psi -
i M_5 {\bar \psi} \gamma_5 \psi
\label{eq:mink}
\end{equation}
becomes, in euclidean space,
\begin{equation}
{\cal L}_{E}=
{\bar \psi} \; (i  \slp+\lnA ) \; \psi+i M {\bar \psi} \psi -
 M_5 {\bar \psi} \gamma_5 \psi.
\label{eq:eucl}
\end{equation}
Notice that the factor of $i$  in the mass term results from our choice
of hermitean euclidean $\gamma$-matrices. Had we chosen antihermitean
$\gamma$-matrices the factor of $i$ would have been absent from the
fermion mass term.

Finally we note that the Minkowski space fermion propagator
\begin{equation}
S_M^{\alpha \beta}= -i \langle {\hat T} \psi_\alpha (x) \; {\bar
\psi}_\beta (x') \rangle
\label{eq:diracpropmink}
\end{equation}
(where $\langle A \rangle$ is the ground state expectation value of $A$
and ${\hat T}$ is the time ordering operator) becomes, in euclidean
space, \begin{equation}
S_E^{\alpha \beta}=  \langle  \psi_\alpha (x) \; {\bar
\psi}_\beta (x') \rangle.
\label{eq:diracpropeucl}
\end{equation}


\newpage
\begin{references}

\bibitem{lieb} D.~C.~Mattis and E.~Lieb, {\sl J.~Math.~Phys.\/}{\bf 6}
(1965) 304.
\bibitem{luther} A.~Luther and V.~J.~Emery, {\sl Phys.~Rev.~Lett.\/}
{\bf 33\/} (1974) 589; V.~J.~Emery, in {\sl Highly Conducting One-Dimensional
Solids}, edited by
J.~DeVreese, R.~Evrard and V.~van Doren (Plenum, 1979).
\bibitem{coleman} S.~Coleman, {\sl Phys.~Rev.~}{\bf D 11} (1975) 2088.
\bibitem{mandelstam} S.~Mandelstam, {\sl Phys.~Rev.~}{\bf D 11} (1975) 3026.
\bibitem{haldane} F.~D.~M.~Haldane, {\sl J.~Phys.~\/} {\bf C14} (1981) 2585
; F.~D.~M.~Haldane, {\sl Phys.~Rev.~Lett.~\/}{\bf 47} (1981) 1840.
\bibitem{witten} E.~Witten,{\sl Comm.~Math.~Phys.~\/}{\bf 92} (1984) 455.
\bibitem{laplata} R.~E.~Gamboa ~Saravi, M.~A.~Muschietti, F.~A.~Schaposnik
and J.~E.~Solomin, {\sl Ann.~Phys.~\/}{\bf 157} (1984) 330.
\bibitem{polyakov} A.~M.~Polyakov and P.~B.~Wiegmann, {\sl Phys.~Lett.\/}
{\bf B131} (1983) 121.
\bibitem{BPZ} A.~A.~Belavin, A.~M.~Polyakov and A.~B.~Zamolodchikov,
{\sl Nucl.~Phys.~\/}{\bf B241} (1984) 333.
\bibitem{KZ} V.~Knizhnik and A.~B.~Zamolodchikov, {\sl Nucl.~Phys.~\/}
{\bf B247} (1986) 83.
\bibitem{cardy} J.~L.~Cardy, {\sl Nucl.~Phys.~\/}{\bf B240} [FS12] (1984)
514.
\bibitem{affleck} I.~K.~Affleck, {\sl Nucl.~Phys.~\/}{\bf B 265} (1986) 809 .
\bibitem{meir} Y.~Meir, N.~S.~Wingreen and P.~Lee {\sl Phys.~Rev.~Lett.~\/}
{\bf 70} (1993) 2601.
\bibitem{kane} C.~L.~Kane and Matthew ~P.~A.~Fisher, {\sl Phys.~Rev.~\/}
{\bf B46\/} (1992) 15233, and references therein.
\bibitem{callan} C.~G.~Callan, {\sl Nucl.~Phys.~\/}{\bf B212} (1983) 391;
H.~Rubakov, {\sl Nucl.~Phys.~\/}{\bf B203} (1982) 311.
\bibitem{affleckimp} I.~K.~Affleck, UBC Preprint (1993) (UBCTP-93-25) and
references
therein.
\bibitem{Affleck} I.~Affleck and A.~W.~Ludwig {\sl Phys.~Rev.~Lett.~\/}
{\bf 67} (1991) 3160;  A.~W.~Ludwig and I.~Affleck {\sl
Phys.~Rev.~Lett.~\/}{\bf 68} (1992) 1046 ; A.~W.~Ludwig and I.~Affleck
{\sl Phys.~Rev.~\/}{\bf B 48} (1993) 7297.
\bibitem{affleckspin} S.~Eggert. and I.~K.~Affleck, {\sl Phys.~Rev.~\/}
{\bf B 46} (1993) 10866.
\bibitem{enviasdedesarrollo} M.~Fuentes, A.~Lopez and E.~Fradkin, in
preparation.
\bibitem{Forman} E.~Forman, {\sl Invent.~Math~\/}, {\bf 88} (1987) 447.
\bibitem{seeley} R.~T.~Seeley,{\sl Am.~J.~Math.} {\bf 91} (1969) 963.
\bibitem{APS} M.~F.~Atiyah, V.~K.~Patodi and I.~M.~Singer,
{\sl Math.~Proc.~Camb.~Phil.~Soc.} {\bf 89} (1975) 43.
\bibitem{cebo} I.S.Iohvidov, {\sl Hankel and Toeplitz Matrices and Forms}
(Birkhauser, 1982).
\bibitem{falomir} H.~Falomir, M.~A.~Muschietti, E.~M.~Santangelo, {\sl
Phys.~Rev.~} {\bf D 43} (1991) 539.
\bibitem{Anderson} G.~Yuval and P.~W.~Anderson, {\sl Phys.~Rev.~\/}{\bf B1}
(1970) 1522.
\bibitem{Nozieres2} P.~Nozieres and C.~T.de~Dominicis, {\sl Phys.~Rev.}
{\bf 178} (1969) 1097.
\bibitem{Kondo} J.~Kondo, {\sl Prog.~Theor.~Phys.~\/}{\bf 32}(1964) 37.
\bibitem{jefe} E.~Fradkin, L.~P.~Kadanoff, {\sl Nucl.~Phys.} {\bf B170}
(1980) 1.
\bibitem{ceva} L.~P.~Kadanoff and H.~Ceva, {\sl Phys.~Rev.}{\bf B3}
(1971) 3918.
\bibitem{caldeira} A.~Caldeira and A.~J.~Leggett, {\sl Physica} {\bf 121A}
(1983) 587; {\sl Ann.~Phys.~(N.~Y.~)\/} {\bf 149}(1983) 374.
\bibitem{fisherzwerger} M.~P.~A.~Fisher and W.~Zwerger,{\sl Phys.~Rev.~\/}
{\bf B32} (1985) 6190.
\bibitem{ludwig} C.~Callan, I.~Klebanov, A.~W.~W.~Ludwig and J.~Maldacena,
{\sl Nucl.~Phys.~\/}{\bf B422}[FS] (1994) 417.
\bibitem{Tsvelik} A.~Tsvelik, Oxford University prerprint (1995) and private
communication.

\end{references}
\end{document}